\documentclass[11pt,english]{article}
\usepackage[T1]{fontenc}
\usepackage[latin9]{inputenc}
\usepackage{geometry}
\usepackage{hyperref}
\usepackage{cite}

\usepackage[usenames,dvipsnames]{color}
\usepackage{graphicx}
\usepackage{amsmath}
\usepackage{amsfonts}
\usepackage{amssymb}
\usepackage{dsfont}
\usepackage{dcolumn}
\usepackage{bm}
\usepackage{slashed}
\usepackage{pstricks}
\usepackage{slashed}
\usepackage{multirow}
\usepackage{array}
\usepackage{rotating}
\usepackage{xcolor}
\usepackage{epstopdf}
\usepackage{fancybox}
\usepackage{fancyvrb}
\usepackage[normalem]{ulem}
\usepackage{color}
\usepackage{mathdots}
\usepackage{mathrsfs}
\usepackage{multirow}
\usepackage{esint}
\usepackage{multirow,bigdelim}
\allowdisplaybreaks

\newcolumntype{x}[1]{
{\centering}p{#1}}%

\newcommand{\GeV}      {~\mathrm{GeV}}
\newcommand{\TeV}      {~\mathrm{TeV}}

\newcommand{\beqn}{\begin{eqnarray}}
\newcommand{\eeqn}{\end{eqnarray}}
\newcommand{\be}{\begin{equation}}
\newcommand{\ee}{\end{equation}}

\newcommand{\mathsym}[1]{{}}

\def \n34{\tilde{\chi}^{0}_{3,4}}

\def\met100{\slashed{E}_T\geq 100 \GeV}

\newcommand{\gappeq}{\mathrel{\rlap {\raise.5ex\hbox{$>$}}
{\lower.5ex\hbox{$\sim$}}}}
\newcommand{\lappeq}{\mathrel{\rlap{\raise.5ex\hbox{$<$}}
{\lower.5ex\hbox{$\sim$}}}}

\def\met{\slashed{E}_{T}}

\begin{document}

\title{\textbf{Higgs diphoton rate and mass enhancement with vector-like leptons\
and the scale of supersymmetry}}

\author{Wan-Zhe Feng$^{\textrm{a}}$\footnote{vicf@ust.hk}~~and Pran~Nath$^{\textrm{b}}$\footnote{nath@neu.edu}}
\date{}
\maketitle
\centerline{$^{\rm a}$ \textit{Department of Physics and Institute for Advanced Study,}}
\centerline{\textit{The Hong Kong University of Science and Technology, Hong Kong}}
\centerline{$^{\rm b}$ \textit{Department of Physics, Northeastern University, Boston, MA 02115, USA}}
\bigskip
\bigskip
\bigskip
\bigskip
\bigskip
\bigskip
\begin{abstract}
Analysis of contributions from vector-like leptonic supermultiplets
to the Higgs diphoton decay rate and to the Higgs boson mass is given.
Corrections arising from the exchange of the new leptons and their super-partners,
as well as their mirrors are computed analytically and numerically.
We also study the correlation between the enhanced Higgs diphoton rate and the Higgs mass corrections.
Specifically, we find two branches in the numerical analysis:
on the lower branch the diphoton rate enhancement is flat
while on the upper branch it has a strong correlation with the Higgs mass enhancement.
It is seen that a factor of 1.4-1.8 enhancement of the Higgs diphoton rate on the upper branch can be achieved,
and a 4-10 GeV positive correction to the Higgs mass can also be obtained simultaneously.
The effect of this extra contribution to the Higgs mass is to release the constraint
on weak scale supersymmetry, allowing its scale
to be lower than in the theory without extra contributions.
The vector-like supermultiplets also have collider implications
which could be testable at the LHC and at the ILC.
\end{abstract}

\newpage

\section{Introduction}

Recently the ATLAS and the CMS Collaborations
using the combined 7~TeV and 8~TeV data found a signal for a boson with the
ATLAS finding a signal at $126.0 \pm0.4 ({\rm stat})\pm 0.4({\rm sys})~{\rm GeV}$ at the $5.0\sigma$ level~\cite{:2012gk}
and the CMS finding a signal at $125.3\pm 0.4 ({\rm stat})\pm 0.5({\rm sys})~{\rm GeV}$ at the $5.0\sigma$ level~\cite{:2012gu}.
While the properties of this boson still need to be fully established there is
the general belief that it is indeed the
long sought after Higgs boson~\cite{Englert:1964et,Higgs:1964pj,Guralnik:1964eu}
of the electroweak theory~\cite{Weinberg:1967tq,salam}.
 In the analysis below we will assume that
the observed  boson is indeed the Higgs particle that is remnant of the electroweak symmetry breaking.
It is pertinent to observe that the results of the ATLAS and CMS Collaborations are remarkably
consistent with the predictions of supergravity grand unified
models~\cite{Chamseddine:1982jx,Nath:1983aw,Hall:1983iz,Arnowitt:1992aq}
with radiative electroweak symmetry breaking (for a review see~\cite{Ibanez:2007pf})
which predict the Higgs boson
mass to lie below around $130$ GeV~\cite{Akula:2011aa,Akula:2012kk,Arbey:2012dq,Ellis:2012aa,Baer:2012mv}
(For a recent review of Higgs and supersymmetry see~\cite{Nath:2012nh}).
However, the fact that the Higgs mass lies close to the
upper limit of the prediction of the  supergravity unification within the
Minimal Supersymmetric Standard Model (MSSM) indicates that the loop  correction to the
Higgs boson mass is rather large which in turn implies that the existence of a high scale of supersymmetry,
specifically a high scale for the squarks. However, corrections on the order of a few GeVs from a
source external to MSSM can significantly lower the scale of supersymmetry. Here we investigate this
possibility by considering an extension of MSSM with vector-like leptonic supermultiplets.
The assumption of additional vector-like leptonic supermultiplets will not alter the Higgs production cross
section and is not strongly constrained by the electroweak data. \\

Aside from the relative heaviness of the Higgs boson is the issue of any possible deviations of
the Higgs boson couplings from the ones predicted in the Standard Model.
If a significant deviation from the Standard Model prediction is seen, it would indicate the existence of new physics.
However, it would take a considerable amount of luminosity, i.e., as much as 3000 fb$^{-1}$ at LHC14 to
achieve an accuracy of 10-20\%~\cite{Peskin:2012we} in the determination of the Higgs couplings with fermions and
with dibosons. An exception to the above is the diphoton channel for which the background is remarkably small
and it  was the discovery channel for the Higgs boson.
Here the current data gives some hint of a possible deviation from the Standard Model prediction.
The ATLAS and the CMS Collaborations give~\cite{:2012gu,:2012gk}:
\beqn
R_{\gamma \gamma}
\equiv
\frac{\sigma(pp\to h)_{\rm obs}}{\sigma(pp\to h)_{\rm SM}}
\cdot \frac{\Gamma(h\to\gamma\gamma)_{\rm obs}}{\Gamma(h\to \gamma\gamma)_{\rm SM}}
= 1.8\pm  0.5 ~({\rm ATLAS }), ~1.6\pm 0.4~({\rm CMS}),
\label{RGG}
\eeqn
where
\beqn
\frac{\sigma(pp\to h)_{\rm obs}}{\sigma(pp\to h)_{\rm SM}}
= 1.4\pm 0.3 {\rm ~(ATLAS)}, ~0.87 \pm 0.23 {\rm ~(CMS)}.
\eeqn
In the Standard Model the largest contribution to the $h\to\gamma\gamma$ mode arises from the $W^+W^-$ in the loop
and this contribution is partially cancelled by the contribution arising from the top quark in the loop.
If this observed enhancement is not due to QCD uncertainties~\cite{Baglio:2012fu}, one needs
new contributions beyond the standard model to increase the diphoton rate.  There are many works which have investigated this possibility, and an enhancement of the diphoton rate can be
achieved in many ways: from light staus with large mixing~\cite{Carena:2011aa,Giudice:2012pf,Sato:2012bf,Basso:2012tr},
from extra vector-like leptons~\cite{Carena:2012xa,An:2012vp,Joglekar:2012vc,ArkaniHamed:2012kq,Almeida:2012bq,Davoudiasl:2012ig,Davoudiasl:2012tu}
and through other mechanisms~\cite{Wang:2012zv,Draper:2012xt,Abe:2012fb,Haba:2012zt,Delgado:2012sm,SchmidtHoberg:2012yy,Urbano:2012tx,Moreau:2012da,
Chala:2012af,Picek:2012ei,Dawson:2012mk,Choi:2012he,SchmidtHoberg:2012ip,Huo:2012tw,Cheung:2012pq,Basso:2012nh}.
Additional papers where vector-like fermions have been discussed
are~\cite{Dawson:2012di,Bonne:2012im,Kearney:2012zi,Voloshin:2012tv,Carmona:2013cq}.
Most of these works are within non-supersymmetric framework. However, with 125 GeV Higgs mass,
vacuum stability is a serious problem in most models. Thus, for example, in the Standard Model vacuum stability
up to the Planck scale may not be achievable since analysis using next-to-next-leading order correction
require that $m_h>129.4$ GeV for the vacuum to be absolutely stable
up to the Planck scale~\cite{Degrassi:2012ry} (see, however, \cite{Alekhin:2012py,Masina:2012tz}).
For this reason we consider supersymmetric models which are less problematic with regard to vacuum
stability (see e.g.,~\cite{ArkaniHamed:2012kq,Hisano:2010re,Kitahara:2012pb,Carena:2012mw}).
Additionally, the supersymmetric theories also avoid the well-known fine-tuning problems
of non-supersymmetric theories.
An analysis to determine whether a significant diphoton enhancement can be achieved in MSSM
was carried out in~\cite{Desai:2012qy,Cao:2013ur}.
\\

In this work, we consider effects from additional vector-like leptonic multiplets in loops both
to the Higgs diphoton rate and to the Higgs mass in a supersymmetric framework.
Vector-like multiplets appear in a variety of
grand unified models~\cite{Georgi:1979md,Wilczek:1981iz,Babu:2005gx} as well as in string and brane models.
Higgs mass enhancement via vector-like supermultiplets has been considered in previous works,
see, e.g.,~\cite{Babu:2008ge,Martin:2009bg,Martin:2010dc}.
New particles with couplings to the Higgs are constrained by  the
electroweak precision tests and such constraints have been discussed
in~\cite{Cynolter:2008ea,Joglekar:2012vc,Almeida:2012bq,ArkaniHamed:2012kq}
and the detection of such particles was discussed in~\cite{Giddings:2013gh}.
The outline of the rest of the paper is as follows: In Section~\ref{Sec2} we give a general analysis of the diphoton
rate in the Standard Model as well as in supersymmetric extensions.  In Section~\ref{Sec3} we discuss the details of the
model. In Section~\ref{Sec4} we give an analysis of the enhancement of the diphoton rate for the model
discussed in the previous section. In Section~\ref{Sec5} we give an analysis of the correction to the Higgs boson mass
from radiative corrections arising from the exchange of the vector-like supermultiplets. A numerical
analysis of the corrections to the Higgs diphoton rate and to the Higgs boson mass is given in Section~\ref{Sec6}
and conclusions are given in Section~\ref{Sec7}.
Further details  are given in Appendix~\ref{AppA} and \ref{AppB}.

\section{A general analysis of the diphoton rate}\label{Sec2}

We first consider the Standard Model case with the Higgs doublet $H^T=(H^+, H^0)$.
The full  decay width of the Higgs $h$ (where $H^0= (v+ h)/\sqrt 2$
and $v=246\GeV$)
at the one-loop level involving the exchange of spin 1, spin 1/2 and spin 0 particles
in the loops is given by
\begin{equation}
\Gamma(h\to\gamma\gamma)=\frac{\alpha^{2}m_{h}^{3}}{1024\pi^{3}}\Big|\frac{g_{hVV}}{m_{V}^{2}}Q_{V}^{2}A_{1}(\tau_{V})+\frac{2g_{hf{f}}}{m_{f}}N_{c,f}Q_{f}^{2}A_{\frac{1}{2}}(\tau_{f})+\frac{g_{hSS}}{m_{S}^{2}}N_{c,S}Q_{S}^{2}A_{0}(\tau_{S})\Big|^{2}\,,
\label{SMdip}
\end{equation}
where $V,f,S$ denote vectors, fermions, and scalars,
$Q,N$ are their charges and numbers (colors),
$A$'s are the loop functions defined in~\cite{Djouadi:2005gj}
and given in Appendix~\ref{AppA}, and $\tau_i = 4m^2_i/m^2_h$.
The couplings $g_{hVV}$ etc are defined by the interaction Lagrangian so that
\beqn
-\mathcal{L}_{\rm int} =  g_{hVV} hV_{\mu}^+V^{-\mu}+ g_{hff} hf\bar f+ g_{hSS} hS\bar S \,.
\eeqn
For the case of the Standard Model one has $g_{hWW}= g_2 M_{W}$ and $g_{hff}= g_2 m_f\big/(2M_W)$,
where $g_2$ is the $SU(2)$ gauge coupling.
 Here it is easily seen that $g_{hWW}/M_W^2= 2g_{hff}/m_f= 2/v$.
In the standard model, the largest contribution to the diphoton rate
is from the $W$ boson  exchange
and this contribution is partially cancelled by the contribution from the top quark exchange.
Thus for the Standard Model Eq.~\eqref{SMdip}  reduces to
\begin{equation}
\Gamma_{\rm SM}(h\to\gamma\gamma)
\approx \frac{\alpha_{em}^{2}m_{h}^{3}}{256v^2\pi^{3}}\Big|A_{1}(\tau_{W})+N_{c}Q_{t}^{2}A_{\frac{1}{2}}(\tau_{t})\Big|^2
\to \frac{\alpha^{2}_{em}m_{h}^{3}}{256v^2\pi^{3}} |\mathcal{A}_{\rm SM}|^2\,,
\label{SMdipR}
\end{equation}
where $\mathcal{A}_{\rm SM}\approx -6.49$.\\

If the masses of the particles running in the loops which give rise to the decay of the Higgs to diphoton,
are much heavier than the Higgs boson,
the decay of $h\to \gamma\gamma$ is governed by an $h\gamma\gamma$ effective coupling which can
be calculated through the photon self-energy corrections~\cite{Ellis:1975ap,Shifman:1979eb} and
reads
\begin{equation}
\mathcal{L}_{h\gamma\gamma}=\frac{\alpha_{em}}{16\pi}h\Big[\sum_{i}b_{i} Q_i^2 \frac{\partial}{\partial v}\log m_{i}^{2}(v)\Big]F_{\mu\nu}F^{\mu\nu}\,.
\label{Lhgg}
\end{equation}
where $b_{i}$ are:
\begin{align}
\label{2}
b_{1}=-7\,, & \qquad{\rm for\ a\ vector\ boson},\\
\label{3}
b_{\frac{1}{2}}=\tfrac{4}{3}\,, & \qquad{\rm for\ a\ Dirac\ fermion},\\
\label{4}
b_{0}=\tfrac{1}{3}\,, & \qquad{\rm for\ a\ charged\ scalar}.
\end{align}
In the large mass limit, the exact one-loop result of Eq.~\eqref{SMdip} agrees with Eq.~\eqref{Lhgg}.
For relative light particles with mass $m$ running in the loop, $b_i$ receives finite mass corrections to the order of $m^2_h\big/4m^2$.
When there are multiple particles carrying the same electric charge circulating in the loops,
one can write a more general expression by replacing
$\log m_{i}^{2}$ by $\log\big(\det M^{2}\big)\,$, where $M^2$ is the mass-squared matrix of the particles
circulating in the loops.
\\

For MSSM one has two Higgs doublets:
\beqn
H_d = \left(\begin{matrix} H_d^0\cr
                           H_d^-\end{matrix}\right)
= \left(\begin{matrix}
             \frac{1}{\sqrt 2}(v_d+\phi_1)\cr
             H_d^-\end{matrix}\right)\,,\qquad
H_u= \left(\begin{matrix}H_u^+\cr
             H_u^0\end{matrix}\right)
=\left(\begin{matrix}H_u^+ \cr
            \frac{1}{\sqrt 2}(v_u+\phi_2) \end{matrix}\right)\,.
\eeqn
where $v_d$ and $v_u$ are the VEVs of $H^0_d$ and $H^0_u$.
Extension of Eq.~\eqref{Lhgg} to the supersymmetric case is straightforward and we have
\begin{equation}
\mathcal{L}_{h\gamma\gamma}^{{\rm SUSY}}=\frac{\alpha_{em}}{16\pi}h\sum_{i}b_{i}Q_i^2\Big[\cos\alpha\frac{\partial}{\partial v_{u}}\log m_{i}^{2}(v_u)-\sin\alpha\frac{\partial}{\partial v_{d}}\log m_{i}^{2}(v_d)\Big]F_{\mu\nu}F^{\mu\nu}\,\,,
\end{equation}
where $\alpha$ is the mixing angle between the two CP-even Higgs in the MSSM.
Eq.~\eqref{SMdip} is also modified in the supersymmetric case as
we identify the lighter CP-even Higgs with the Standard Model Higgs:
\begin{align}
\Gamma_{\rm SUSY}(h\to\gamma\gamma)&\approx
\frac{\alpha_{em}^{2}m_{h}^{3}}{256 v^2 \pi^{3}}\bigg|
\sin(\beta-\alpha) Q_{W}^{2} A_1(\tau_W)
+\frac{\cos \alpha}{\sin \beta} N_t Q^2_t A_{\frac{1}{2}}(\tau_t)\nonumber\\
&\qquad\qquad\quad
+\frac{b_{\frac{1}{2}} v}{2} N_f Q_{f}^{2} \Big(\cos\alpha \frac{\partial}{\partial v_u} \log m_{f}^{2}
-\sin\alpha \frac{\partial}{\partial v_d} \log m_{f}^{2} \Big) \nonumber\\
&\qquad\qquad\quad
+\frac{b_0 v}{2} N_{c,S} Q_{S}^{2} \Big(\cos\alpha \frac{\partial}{\partial v_u} \log m_{S}^{2}
-\sin\alpha \frac{\partial}{\partial v_d} \log m_{S}^{2} \Big)
\bigg|^{2}\,,\label{SUSYdip}
\end{align}
where $\tan\beta= v_u/v_d$.
Compared to the Standard Model case, the Higgs couplings to the $W$ boson and to the top quark
are modified by factors $\sin(\beta-\alpha)$ and $\frac{\cos \alpha}{\sin \beta}$  (see Eq. \ref{SUSYdip}).
Now the fermionic contribution also comes from the chargino exchange while
the scalar contribution includes contributions from the exchange of  the sleptons, the squarks and the charged Higgs fields.

\section{The Model}\label{Sec3}

To enhance the Higgs diphoton decay rate we focus on the contribution of the vector-like leptonic supermultiplets,
since relatively light vector-like quark supermultiplets would affect the Higgs production cross sections while
leptonic supermultiplets would not.
Specifically we consider an extra vector-like leptonic generation $F$ consisting of $L,L^c,E,E^c$
with $SU(3)_C\times SU(2)_L\times U(1)_Y$ quantum numbers:\footnote{
Gauge coupling unification can be achieved with a full generation of vector-like
multiplets including a vector-like quark sector. We assume relatively large masses and negligible Yukawa couplings
for the quark sector and thus these additions would not contribute  to the diphoton rate or to the Higgs
mass enhancement.}
\begin{equation}
F:\quad \begin{array}{cc}
L=(\mathbf{1},\mathbf{2},-\tfrac{1}{2})\,, & \qquad E^{c}=(\mathbf{1},\mathbf{1},1)\,,\\
L^{c}=(\mathbf{1},\mathbf{2},+\tfrac{1}{2})\,, & \qquad E=(\mathbf{1},\mathbf{1},-1)\,.
\end{array}
\end{equation}
Noting that the Higgs doublets in the MSSM have quantum numbers
\begin{gather}
H_d=(\mathbf{1},\mathbf{2},-\tfrac{1}{2})\,,\qquad H_u=(\mathbf{1},\mathbf{2},+\tfrac{1}{2})\,,
\end{gather}
the superpotential for the vector-like leptonic supermultiplets is given by
\begin{equation}
W = yLH_{d}E^{c}+y'L^{c}H_{u}E+M_{L}LL^{c}+M_{E}EE^{c}+y_{1}^{(m)}L_{3}H_{d}E^{c}+y_{2}^{(m)}LH_{d}E_{3}^{c}\,,
\end{equation}
where $M_{L}$ and $M_{E}$ are the vector-like masses. We assume
that the extra leptons can decay only through the third generation particles,
and the corresponding couplings $y_{1,2}^{(m)}$
are assumed to be very small and they do not have any significant effect on the analysis here.\footnote{
The new leptons could mix with other generations as well.
The reason for allowing for the mixings is to make the new leptons unstable.
This instability can  be accomplished with very small mixing angles, e.g., $\mathcal{O}(10^{-4})$ or even smaller.
Because of this  there is no tangible effect on any analyses involving the three generations of leptons.
There is one area, however,  where LFV could manifest, and that is the  decay of the $\tau'\to \tau + \gamma$
very much like the possibility of the decay $\tau\to \mu + \gamma$~\cite{Ibrahim:2012ds}. The  mixings can also
lead to the edm of the tau lepton~\cite{Ibrahim:2010va} in  the presence of CP phases.}
Neglecting these small terms, the fermionic mass matrix now reduces to
\begin{equation}
M_{F}=\left(\begin{array}{cc}
M_{L} & \tfrac{1}{\sqrt{2}}yv_{d}\\
\tfrac{1}{\sqrt{2}}y'v_{u} & M_{E}
\end{array}\right)\,,
\label{fermimass}
\end{equation}
where the off-diagonal elements are the masses generated by Yukawa interactions while the
diagonal elements are the vector masses.
The two squared-mass eigenvalues arising from Eq.~\eqref{fermimass} are given by
\begin{align}
m^2_{1,2} & =\frac{1}{4}\Big[2M^2_L+2M^2_E+y'^2 v^2_u+y^2 v^2_d \nonumber\\
&\qquad \pm \sqrt{(2M^2_L+2M^2_E+y'^2 v^2_u+y^2 v^2_d)^2-4(2M_L M_E-y y' v_u v_d)^2 }\Big]\,.
\label{m1m2}
\end{align}
We call the heavier one $\tau'_1$ and the lighter one $\tau'_2$.
We note that the neutral component of the $SU(2)$ doublet $L,L^c$
do not play any role in the analysis as they do not enter in the analysis
of the diphoton rate or  in the analysis of the Higgs mass enhancement.

\section{Enhancement of the  diphoton decay rate of the Higgs boson}\label{Sec4}

Inclusion of the vector-like supermultiplet affects the diphoton rate.
Using Eqs.~\eqref{SMdipR} and \eqref{SUSYdip}, the ratio of the decay width of the lighter CP-even
Higgs to two photons and the Standard Model prediction can be written as
\begin{align}
\frac{\Gamma(h\to\gamma\gamma)}{\Gamma(h\to\gamma\gamma)_{{\rm SM}}}
&\approx \frac{1}{|\mathcal{A}_{\rm SM}|^2} \bigg|\sin(\beta-\alpha)Q^2_W A_1(\tau_W)
+\frac{\cos \alpha}{\sin \beta} N_t Q^2_t A_{\frac{1}{2}}(\tau_t)\nonumber\\
&\qquad\qquad\quad
+\frac{b_{\frac{1}{2}} v}{2} N_F Q_{F}^{2} \Big(\cos\alpha \frac{\partial}{\partial v_u} \log M_{F}^{2}
-\sin\alpha \frac{\partial}{\partial v_d} \log M_{F}^{2} \Big) \nonumber\\
&\qquad\qquad\quad
+\frac{b_0 v}{2} N_{S}Q_{S}^{2} \Big(\cos\alpha \frac{\partial}{\partial v_u} \log M_{S}^{2}
-\sin\alpha \frac{\partial}{\partial v_d} \log M_{S}^{2} \Big)
\bigg|^{2}
\,,\label{ratiotoSM}
\end{align}
where on the second line we have fermionic contribution from the vector-like fermions and
on the third line the scalar contribution from the super-partners of the vector-like fermions.
In the analysis here we focus only on the extra contributions arising
from the exchange of the leptonic vector-like sector,
and do  not include other possible corrections to the diphoton rate such as
from the exchange of staus, charginos and charged Higgs in the loops.
\\

The computation of the vector-like fermion contribution is straightforward, and we find
\begin{equation}
\sum_i \big[\cos\alpha\frac{\partial}{\partial v_{u}}\log m^2_{i}-\sin\alpha\frac{\partial}{\partial v_{d}}\log m^2_{i}\big]
=-\frac{yy'v}{m_{1}m_{2}}\cos(\alpha+\beta)\,,
\end{equation}
where
\begin{equation}
m_1 m_2 = M_L M_E - \frac{1}{2} y y' v_u v_d\,.
\end{equation}
For the case when $M_L=M_E=0$, the fermionic contribution to the diphoton rate is negative.
However, for the case when $M_L, M_E\neq 0$ the fermionic contribution can turn positive
when $M_L M_E > \frac{1}{2} y y' v_u v_d$.
If the contribution is only from the vector-like fermions, the Higgs diphoton
rate is enhanced by a factor of:
\begin{align}
\frac{\Gamma(h\to\gamma\gamma)}{\Gamma(h\to\gamma\gamma)_{{\rm SM}}}
& \approx \Big|1+\frac{1}{\mathcal{A}_{{\rm SM}}}b_{\tfrac{1}{2}}N_{f}Q_{f}^{2}\frac{-v^{2}yy'}{2m_{1}m_{2}}\cos(\alpha+\beta)\Big|^{2}\nonumber \\
& \approx\Big|1+0.1 N_f \frac{v^{2}yy'}{m_{1}m_{2}}\cos(\alpha+\beta)\Big|^{2}
\equiv |1+r_f|^2 \,.
 \label{fermionCon}
\end{align}
\\

To determine the contribution from the four super-partner fields of the vector-like fermions,
one needs to find the mass eigenvalues of a
$4\times 4$ mass mixing matrix. In the basis  $(\tilde\tau_L', \tilde\tau_R', \tilde\tau_{L}'', \tilde\tau_R'')$
it is given by
\begin{equation}
\frac{1}{\sqrt{2}}
\left(\begin{array}{cc|cc}
\multicolumn{2}{c|}{\multirow{2}*{$\sqrt{2}(M_{\tilde{\tau}'}^2)_{2\times 2}$}} & y' v_u M_L + y v_d M_E & 0\\
&& 0 & y' v_u M_E + y v_d M_L \\
\hline
y' v_u M_L + y v_d M_E & 0 & \multicolumn{2}{c}{\multirow{2}*{$\sqrt{2} (M_{\tilde{\tau}''}^2)_{2\times 2}$}}\\
0 & y' v_u M_E + y v_d M_L &&\\
\end{array}\right)_{4\times 4}\,,
\label{slep1}
\end{equation}
where $(M^2_{\tilde \tau'})_{2\times 2}$ is given by
\beqn
(M^2_{\tilde \tau'})_{2\times 2}=\left(\begin{array}{cc}
M_{1}^{2}+\tfrac{1}{2}y^2 v^2_d +M_{L}^{2} +\frac{(g_1^2-g_2^2)}{8} (v_d^2 - v_u^2)
& \frac{1}{\sqrt 2} y (A_{\tau'} v_d  - \mu v_u)\\
\frac{1}{\sqrt 2} y (A_{\tau'} v_d  - \mu v_u)
&   M_1^2 +\tfrac{1}{2}y^2 v^2_d+ M_E^2 - \frac{g_1^2}{4} (v_d^2 - v_u^2)
\end{array}\right)\,,
\label{slep2}
\eeqn
and  $(M^2_{\tilde \tau''})_{2\times 2}$ is given by
\beqn
(M^2_{\tilde \tau''})_{2\times 2}
=\left(\begin{array}{cc}
M_{2}^{2}+\tfrac{1}{2}y'^2 v^2_u+M_{L}^{2} -\frac{(g_1^2-g_2^2)}{8} (v_d^2 - v_u^2)
& \frac{1}{\sqrt 2} y' (A_{\tau''} v_u  - \mu v_d)\\
\frac{1}{\sqrt 2} y' (A_{\tau''} v_u  - \mu v_d)
&   M_2^2 +\tfrac{1}{2}y'^2 v^2_u+ M_E^2 +\frac{g_1^2}{4} (v_d^2 - v_u^2)
\end{array}\right)\,,
\label{slep3}
\eeqn
where $M_1, M_2$ are soft scalar masses.
For further convenience, we define $M^2 \equiv M_{\tilde{\tau}'}^2$ and $M'^2 \equiv M_{\tilde{\tau}''}^2$.
As an approximation, we consider the case when the soft squared-mass ($M_{1,2}^2$) are  much larger
than the vector squared-mass ($M_{L,E}^2$). In this case, the $4\times4$ matrix becomes approximately block diagonal
with the diagonal elements consisting of two $2\times 2$ matrices.
As the two mass-squared matrices are decoupled, we  denote the
super-partners of $\tau'$ to be $\tilde{\tau}'_{1,2}$
and the super-partners of $\tau''$ to be $\tilde{\tau}''_{1,2}$.
The contributions from the two decoupled matrices can be obtained straightforwardly.
The total bosonic contribution can be measured by $r_b$, which reads
\begin{equation}
r_b = r_1 + r_2 \equiv \frac{1}{\mathcal{A}_{\rm SM}} \frac{b_0 v}{2} Q_S^2 (\Xi_{1} + \Xi_2)\,,
\label{rb}
\end{equation}
where we define
\begin{align}
\Xi_1 &= \cos \alpha \frac{\partial}{\partial v_{u}} \log (\det M^2) - \sin\alpha  \frac{\partial}{\partial v_{d}} \log (\det M^2)\,,\\
\Xi_2 &= \cos \alpha \frac{\partial}{\partial v_{u}} \log (\det M'^2) - \sin\alpha  \frac{\partial}{\partial v_{d}} \log (\det M'^2)\,.
\end{align}
We first focus on $\Xi_1$. Using the $\tilde\tau'$ mass-squared matrix, a direct computation gives
\beqn
\Xi_1 =
\frac{1}{ m_{\tilde\tau_1'}^2 m_{\tilde \tau_2'}^2}
\left\{
\Big[ \frac{1}{2} g_1^2 M_{11}^2 - \frac{(g_1^2-g_2^2)}{4} M_{22}^2\Big] v \sin(\alpha+\beta)
 +\sqrt 2 M_{12}^2 y (A_{\tau'} \sin\alpha + \mu \cos\alpha ) \right\} \,.
 \label{xi1}
\eeqn
For the computation of $\Xi_2$ we need  $\tilde{\tau}''$ mass-squared matrix and a similar
analysis gives
\beqn
\Xi_2 =
\frac{1}{ m_{\tilde\tau_1''}^2 m_{\tilde \tau_2''}^2}
\left\{
\Big[
-\frac{1}{2} g_1^2 M'^2_{11} + \frac{(g_1^2-g_2^2)}{4} M'^2_{22}
\Big] v \sin(\alpha+\beta)
- \sqrt 2 M'^2_{12} y' (A_{\tau''} \cos\alpha + \mu \sin\alpha) \right\} \,.
\label{xi2}
\eeqn
\\

Thus the total Higgs diphoton decay rate is enhanced by a factor
\beqn
R_{\gamma\gamma}=\big|1+ r_f + r_b\big|^2\,.
\label{RGG2}
\eeqn
A numerical analysis of the size of the diphoton rate enhancement using the result of this
section is discussed in Section~\ref{Sec6}.
For the numerical analysis, we made the same approximation as above, i.e.,
we choose the value of soft squared-mass to be much larger than the value of vector squared-mass.

\section{Higgs mass enhancement}\label{Sec5}

Extra  particles beyond those in MSSM can make contributions to the mass of the Higgs boson.
In our case, contributions arise from the exchange of both bosonic and fermionic particles
in the vector-like supermultiplets.
The techniques for the computation of these corrections are well-known (see, e.g., \cite{Ibrahim:2000qj,Ibrahim:2002zk})
and are described in Appendix~\ref{AppB}.
Effectively the corrections are encoded in elements $\Delta_{ij}$ which are corrections to
the elements of tree-level mass-squared matrix as defined also in the Appendix~\ref{AppB}.
The correction to the lighter CP-even Higgs mass is then given by
\beqn
(\Delta m_h)_{F}=(2m_h^0)^{-1}
(\Delta_{11 } \sin^2\alpha
+\Delta_{22} \cos^2\alpha - \Delta_{12} \sin2\alpha)\,,
\label{HiggsMassCor}
\eeqn
where $\alpha$ is the mixing angle between the two CP-even Higgs in the MSSM.
Thus, one can write the Higgs boson mass in the form
\begin{equation}
m_h=  m_h^{\rm MSSM} + (\Delta m_h)_{F}\,,
\end{equation}
where $m_h^{\rm MSSM}$ is the Higgs boson mass in the MSSM and $(\Delta m_h)_{F}$ is the correction
from the new sector given by Eq.~\eqref{HiggsMassCor}.
In the following we will first discuss the contribution to the lightest Higgs boson mass from the
bosonic sector and then from the fermionic sector of the vector-like supermultiplets.
The total contribution to the Higgs mass is the sum of bosonic and fermionic contributions, and we have
\beqn
\Delta_{ij} = \Delta_{ij}^b + \Delta_{ij}^f\,.   \label{deltaij}
\eeqn
We note that the coupling between the $\tau'$ and the $\tau''$ sectors are characterized by $M_L$ and $M_E$.
For the case when $M_L=M_E=0$ the $\tau'$ and the $\tau''$ sectors
(both bosonic and fermionic sectors) totally decouple.
In this circumstance one can calculate $\Delta_{ij}$ analytically.

\subsection{Higgs mass correction from the bosonic sector}

The mass-squared matrix in the bosonic sector is given by Eqs.~\eqref{slep1}-\eqref{slep3}.
Here again, we choose the soft squared-mass to be much larger than $M_L^2$ and $M_E^2$.
In this circumstance the $4\times 4$ mass-squared matrix of Eq.~\eqref{slep1} becomes approximately
block diagonal and one can obtain the results for
Higgs mass enhancement from the super-partners of the vector-like fermions ($\tilde{\tau}'_{1,2}$ and $\tilde{\tau}''_{1,2}$).
We first compute the corrections from $\tilde{\tau}'_{1,2}$. The computation of the corrections
uses the Coleman-Weinberg one-loop effective potential~\cite{Coleman:1973jx,Arnowitt:1992qp}
(see Appendix~\ref{AppB}).
The contribution to this one-loop effective potential from $\tilde{\tau}'_{1,2}$ exchanges is given by
\beqn
\Delta V_{\tilde{\tau}'}^b=\frac{1}{64\pi^2}
\sum_{i=1,2} 2 m_{\tilde{\tau}'_i}^4 \Big(\ln \frac{m_{\tilde{\tau}'_i}^2}{Q^2}-\frac{3}{2}\Big)
\,,
\eeqn
where $Q$ is the running renormalization group  scale.
Our computation of $\Delta_{ij}^{\tilde{\tau}'}$  following the  prescription
 in Appendix~\ref{AppB} (further details can be found in~\cite{Ibrahim:2000qj,Ibrahim:2002zk})
gives
\begin{align}
\Delta_{11}^{\tilde{\tau}'}
&=
\beta y^4 v^2_d
\ln \frac{m_{\tilde{\tau}'_1}^2 m_{\tilde{\tau}'_2}^2}{Q^4}
-\beta y^4 v^2_d A_{\tau'}^2
\frac{(A_{\tau'} -\mu\tan\beta)^2}
{(m_{\tilde{\tau}'_1}^2-m_{\tilde{\tau}'_2}^2)^2}
f(m_{\tilde{\tau}'_1}^2, m_{\tilde{\tau}'_2}^2) \nonumber\\
&\,\quad + 2 \beta y^4 v^2_d A_{\tau'}
\frac{A_{\tau'} -\mu\tan\beta}
{m_{\tilde{\tau}'_1}^2-m_{\tilde{\tau}'_2}^2}
\ln \frac{m_{\tilde{\tau}'_1}^2}{m_{\tilde{\tau}'_2}^2} \,,\label{Db'11}\\
\Delta_{22}^{\tilde{\tau}'}
&= - \beta y^4 v^2_d \mu^2
\frac{(A_{\tau'}-\mu\tan\beta)^2}
{(m_{\tilde{\tau}'_1}^2-m_{\tilde{\tau}'_2}^2)^2}
f(m_{\tilde{\tau}'_1}^2, m_{\tilde{\tau}'_2}^2)\,,\\
\Delta_{12}^{\tilde{\tau}'}
&=
\beta y^4 v^2_d \mu A_{\tau'}
\frac{(A_{\tau'} -\mu\tan\beta)^2}
{(m_{\tilde{\tau}'_1}^2-m_{\tilde{\tau}'_2}^2)^2}
f(m_{\tilde{\tau}'_1}^2, m_{\tilde{\tau}'_2}^2)
- \beta y^4 v^2_d \mu
\frac{A_{\tau'} -\mu\tan\beta}
{m_{\tilde{\tau}'_1}^2-m_{\tilde{\tau}'_2}^2}
\ln \frac{m_{\tilde{\tau}'_1}^2}{m_{\tilde{\tau}'_2}^2}\,,\label{Db'12}
\end{align}
where $\beta=1/16\pi^2$,
and $f(x,y)$ is given by
\beqn
f(x,y)= -2 + \frac{y+x}{y-x} \ln \frac{y}{x}\,.
\eeqn
The contribution to the one-loop effective potential from $\tilde{\tau}''_{1,2}$  exchanges is given by
\beqn
\Delta V_{\tilde{\tau}''}^b=\frac{1}{64\pi^2}
\sum_{i=1,2} 2 m_{\tilde{\tau}''_i}^4\Big(\ln \frac{m_{\tilde{\tau}''_i}^2}{Q^2}-\frac{3}{2}\Big)
\,.
\eeqn
A similar computation gives the result for $\Delta_{ij}^{\tilde{\tau}''}$:
\begin{align}
\Delta_{11}^{\tilde{\tau}''}
&=- \beta y'^4 v_u^2 \mu^2
\frac{(A_{\tau''}-\mu\cot\beta)^2}
{(m_{\tilde{\tau}''_1}^2-m_{\tilde{\tau}''_2}^2)^2}
f(m_{\tilde{\tau}''_1}^2, m_{\tilde{\tau}''_2}^2)\,,\label{Db''11}\\
\Delta_{22}^{\tilde{\tau}''}
&=
 \beta y'^4 v_u^2
\ln \frac{m_{\tilde{\tau}''_1}^2 m_{\tilde{\tau}''_2}^2}{Q^4}
- \beta y'^4 v_u^2 A_{\tau''}^2
\frac{(A_{\tau''} -\mu\cot\beta)^2}
{(m_{\tilde{\tau}''_1}^2-m_{\tilde{\tau}''_2}^2)^2}
f(m_{\tilde{\tau}''_1}^2, m_{\tilde{\tau}''_2}^2) \nonumber\\
&
\,\quad + 2 \beta y'^4 v_u^2 A_{\tau''}
\frac{A_{\tau''} -\mu\cot\beta}
{m_{\tilde{\tau}''_1}^2-m_{\tilde{\tau}''_2}^2}
\ln \frac{m_{\tilde{\tau}''_1}^2}{m_{\tilde{\tau}''_2}^2}\,, \label{Db''22}\\
\Delta_{12}^{\tilde{\tau}''}
&=-\beta y'^4 v_u^2 \mu
\frac{A_{\tau''} -\mu\cot\beta}
{m_{\tilde{\tau}''_1}^2-m_{\tilde{\tau}''_2}^2}
\ln \frac{m_{\tilde{\tau}''_1}^2}{m_{\tilde{\tau}''_2}^2}
+ \beta y'^4 v_u^2 \mu A_{\tau''}
\frac{(A_{\tau''} -\mu\cot\beta)^2}
{(m_{\tilde{\tau}''_1}^2-m_{\tilde{\tau}''_2}^2)^2}
f(m_{\tilde{\tau}''_1}^2, m_{\tilde{\tau}''_2}^2)\,.\label{Db''12}
\end{align}
The total contribution from the bosonic sector of the vector-like supermultiplets $\Delta_{ij}^b$ is then
\beqn
\Delta_{ij}^b= \Delta_{ij}^{\tilde{\tau}'} + \Delta_{ij}^{\tilde{\tau}''}\,.
\label{bosonicdelta}
\eeqn

\subsection{Corrections to the Higgs boson mass from the fermionic sector}

We now turn to a discussion of the contribution from the fermionic sector of the vector-like supermultiplet.
Here, in contrast to the bosonic sector, there are no soft terms and further the vector masses can be
comparable to the masses arising from Yukawa couplings. As a consequence  $M_L, M_E$ should be
included in the analysis for a reliable estimate of the contribution from the fermionic sector to the
Higgs mass correction. The contribution to the one-loop effective potential from the vector-like fermions
is given by
\begin{equation}
\Delta V_{\tau'_{1,2}}^f=
-\frac{1}{64\pi^2}
\sum_{i=1,2} 4 m_{i}^4 \Big(\ln\frac{m_{i}^2}{Q^2}-\frac{3}{2}\Big)\,,
\end{equation}
where $m_{1,2}$ are the mass eigenvalues of the vector-like fermions which are given in Eq.~\eqref{m1m2}.
A straightforward analysis  gives
\begin{align}
\Delta_{11}^f & =-\beta \Big[\big(\frac{1}{2}y^{4}v_{d}^{2}-\frac{1}{2}N_{1}\sqrt{R}+\frac{R_{d}^{\prime2}}{8R}\big)\ln\frac{m_{1}^{2}m_{2}^{2}}{Q^{4}}+\big(\frac{y^{2}v_{d}R'_{d}}{2\sqrt{R}}-\frac{1}{2}N_{1}T\big)\ln\frac{m_{1}^{2}}{m_{2}^{2}}+N_{1}\sqrt{R}\Big]\,,
\label{Df11nz}\\
\Delta_{22}^f & =-\beta \Big[\big(\frac{1}{2}y'^{4}v_{u}^{2}-\frac{1}{2}N_{2}\sqrt{R}+\frac{R_{u}^{\prime2}}{8R}\big)\ln\frac{m_{1}^{2}m_{2}^{2}}{Q^{4}}+\big(\frac{y'^{2}v_{u}R'_{u}}{2\sqrt{R}}-\frac{1}{2}N_{2}T\big)\ln\frac{m_{1}^{2}}{m_{2}^{2}}+N_{2}\sqrt{R}\Big]\,,
\label{Df22nz}\\
\Delta_{12}^f & =-\beta \Big[\big(\frac{1}{2}y^{2}y'^{2}v_{u}v_{d}-\frac{1}{2}N\sqrt{R}+\frac{R'_{u}R'_{d}}{8R}\big)\ln\frac{m_{1}^{2}m_{2}^{2}}{Q^{4}}\nonumber\\
 & \qquad\qquad\qquad\qquad\qquad\qquad\qquad\quad+\big(\frac{y^{2}v_{d}R'_{u}+y'^{2}v_{u}R'_{d}}{4\sqrt{R}}-\frac{1}{2}NT\big)\ln\frac{m_{1}^{2}}{m_{2}^{2}}+N\sqrt{R}\Big]\,,
 \label{Df12nz}
\end{align}
where
\begin{align}
T & =M_{L}^{2}+M_{E}^{2}+\frac{1}{2}y'^{2}v_{u}^{2}+\frac{1}{2}y^{2}v_{d}^{2}\,,\\
R & =T^{2}-(2M_{L}M_{E}-yy'v_{u}v_{d})^{2}\,,\\
N_{1} & =\frac{R'_{d}}{2v_{d}\sqrt{R}}+\frac{R_{d}^{\prime2}}{4\sqrt{R^{3}}}-\frac{R''_{d}}{2\sqrt{R}}\,,\\
N_{2} & =\frac{R'_{u}}{2v_{u}\sqrt{R}}+\frac{R_{u}^{\prime2}}{4\sqrt{R^{3}}}-\frac{R''_{u}}{2\sqrt{R}}\,,\\
N & =\frac{R'_{u}R'_{d}}{4\sqrt{R^{3}}}-\frac{R''_{ud}}{2\sqrt{R}}\,,
\end{align}
and
\begin{gather}
R'_{d}=\frac{\partial R}{\partial v_{d}}\,,\qquad R'_{u}=\frac{\partial R}{\partial v_{u}}\,,\\
R''_{d}=\frac{\partial^{2}R}{\partial v_{d}^{2}}\,,\qquad R'_{u}=\frac{\partial^{2}R}{\partial v_{u}^{2}}\,,\qquad R''_{ud}=\frac{\partial^{2}R}{\partial v_{u}\partial v_{d}}\,.
\end{gather}
As a check, we consider the limit  $M_L=M_E=0$. In this limit
$m_1\to \frac{1}{\sqrt{2}} y' v_u$ and
$m_2\to \frac{1}{\sqrt{2}} y v_d$  and Eqs.~\eqref{Df11nz}-\eqref{Df12nz} reduce to
\begin{align}
\Delta_{11}^f & =- \beta y^4 v_d^2 \ln \frac{y^4 v_d^4}{4Q^{4}}\,, \label{Df11z}\\
\Delta_{22}^f & =- \beta y'^4 v_u^2 \ln\frac{y'^4 v_u^2}{4Q^{4}}\,, \label{Df22z}\\
\Delta_{12}^f & =0\,. \label{Df12z}
\end{align}
These are precisely the results that we expect in the decoupled limit.
In this limit, combining Eqs.~\ref{Db'11} with \ref{Df11z}, and Eqs.~\ref{Db''22} with \ref{Df22z},
we find that the $Q$ dependence cancels out and the entire one loop correction is independent
of $Q$.
For the case that $M_L,M_E\neq 0$, one  also expects that the $Q$ dependence would drop out
when we combine the bosonic and the fermionic contributions.
However, the analysis in the bosonic sector is only approximate, and thus there may be a
small residual $Q$ dependence in the total bosonic and fermionic contribution.
However, in the next section we check on the $Q$ dependence numerically and show that
the $Q$ dependence of the total bosonic and fermionic contribution is extremely small validating
our approximation in the computation of the bosonic contribution.

\section{Numerical analysis}\label{Sec6}

\begin{figure}[t!]
\begin{center}
\includegraphics[scale=0.33]{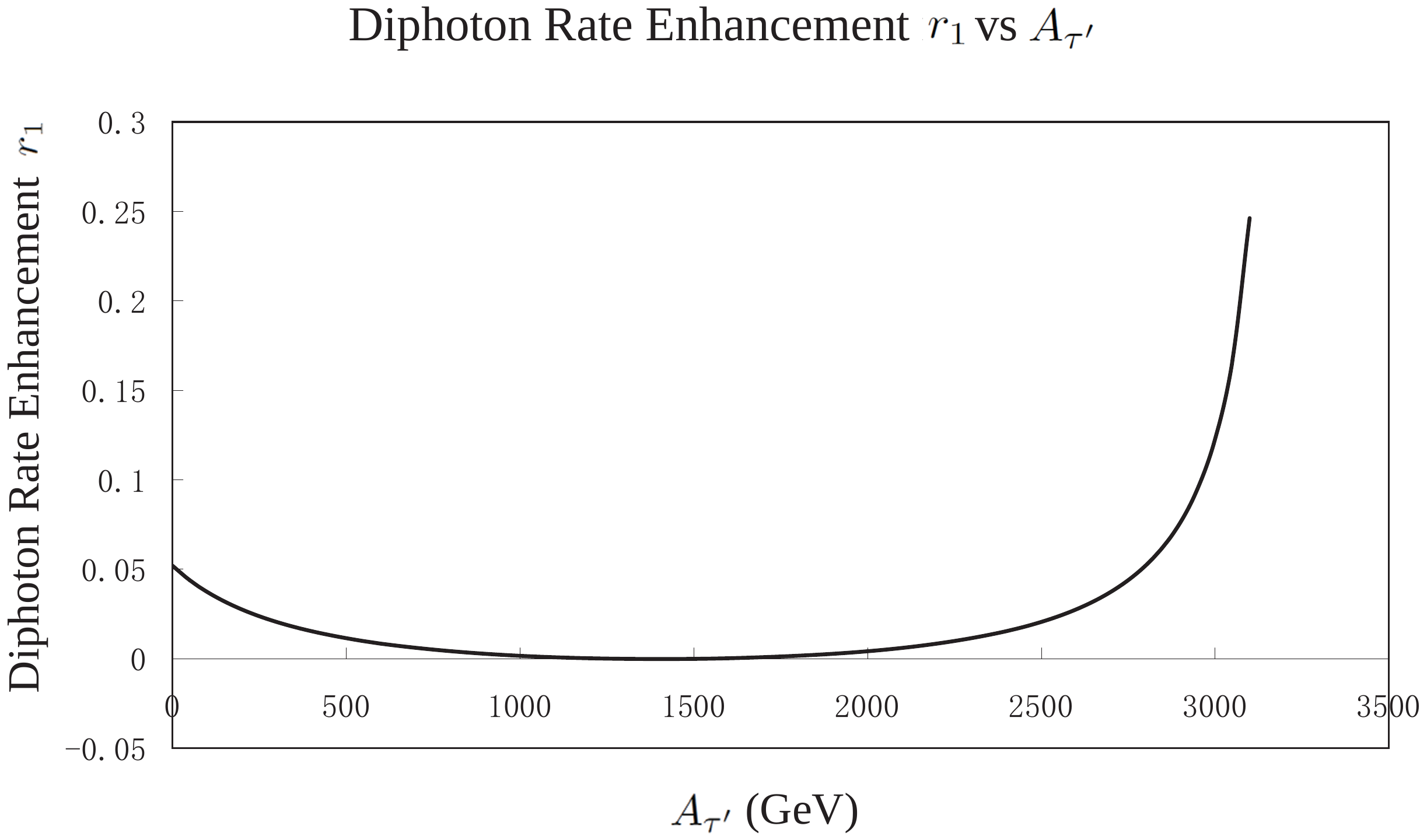}
\includegraphics[scale=0.33]{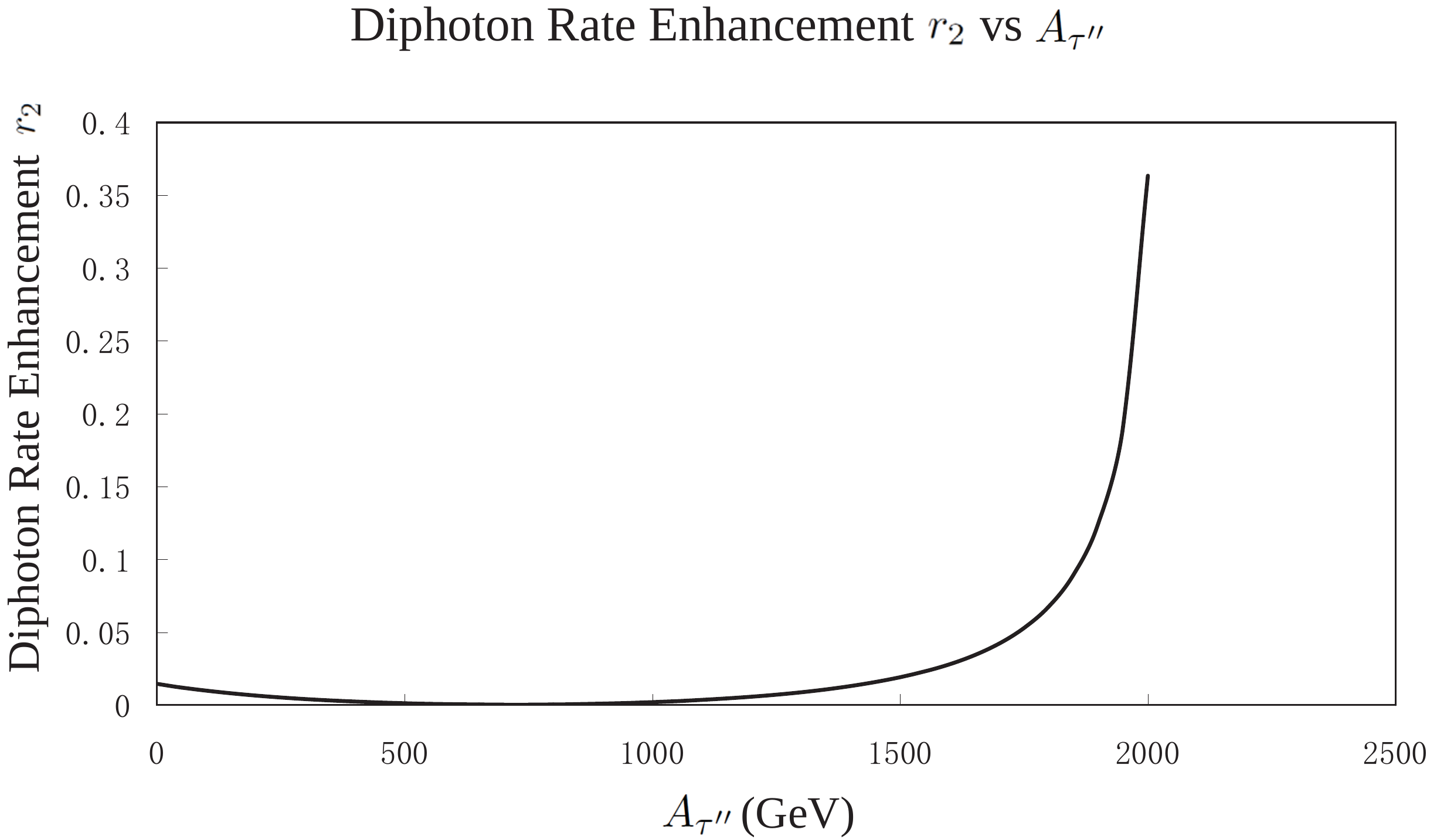}
\\~\\
\includegraphics[scale=0.33]{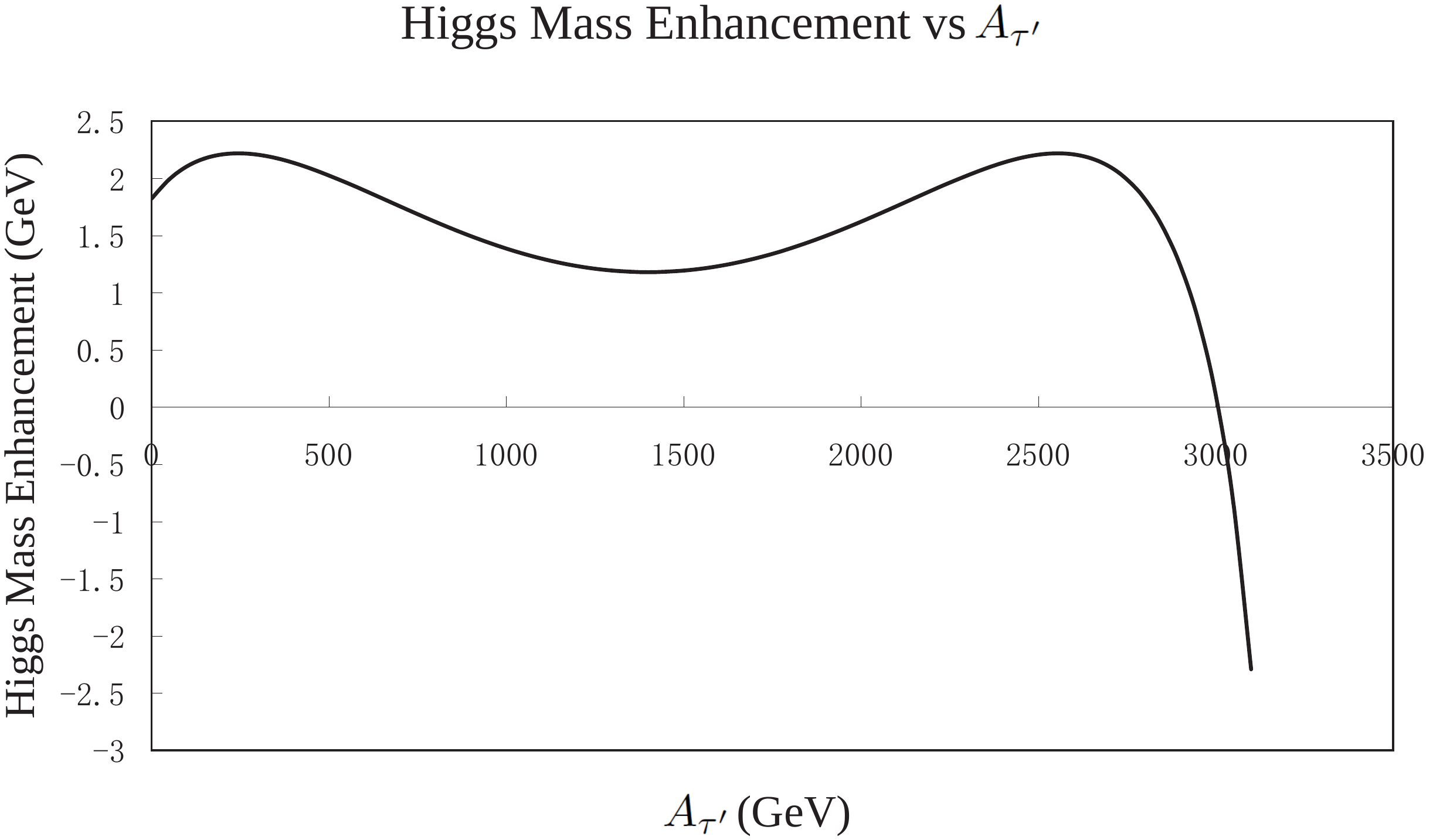}
~\includegraphics[scale=0.325]{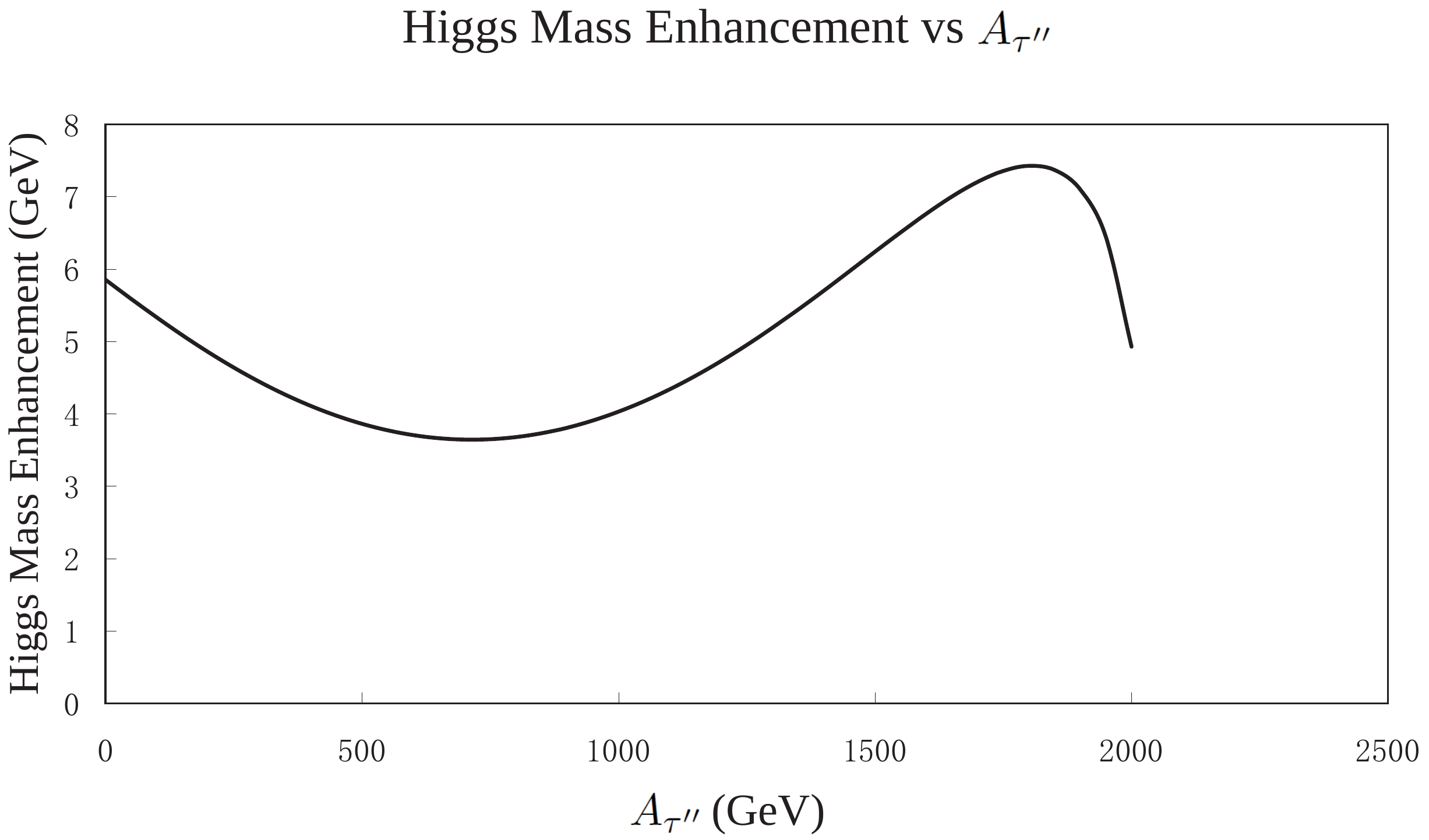}
\caption{An analysis of the diphoton rate  enhancement (top panels) and enhancement of the Higgs boson mass
(bottom panels) for the case when the vector masses vanish, i.e., $M_L=M_E=0$.
Left top: A plot of the diphoton rate enhancement $r_1$ (from $\tilde{\tau}'_{1,2}$) vs  $A_{\tau'}$;
Right top: A  plot of the diphoton rate enhancement $r_2$ (from $\tilde{\tau}''_{1,2}$) vs $A_{\tau''}$.
Left bottom: A plot of the Higgs mass enhancement from $\hat{\tau}'$ sector (GeV) vs $A_{\tau'}$;
Right bottom:  A plot of the Higgs mass enhancement from $\hat{\tau}''$ sector (GeV) vs $A_{\tau''}$.
}
\label{fig1}
\end{center}
\end{figure}
\begin{figure}[t!]
\begin{center}
\includegraphics[scale=0.33]{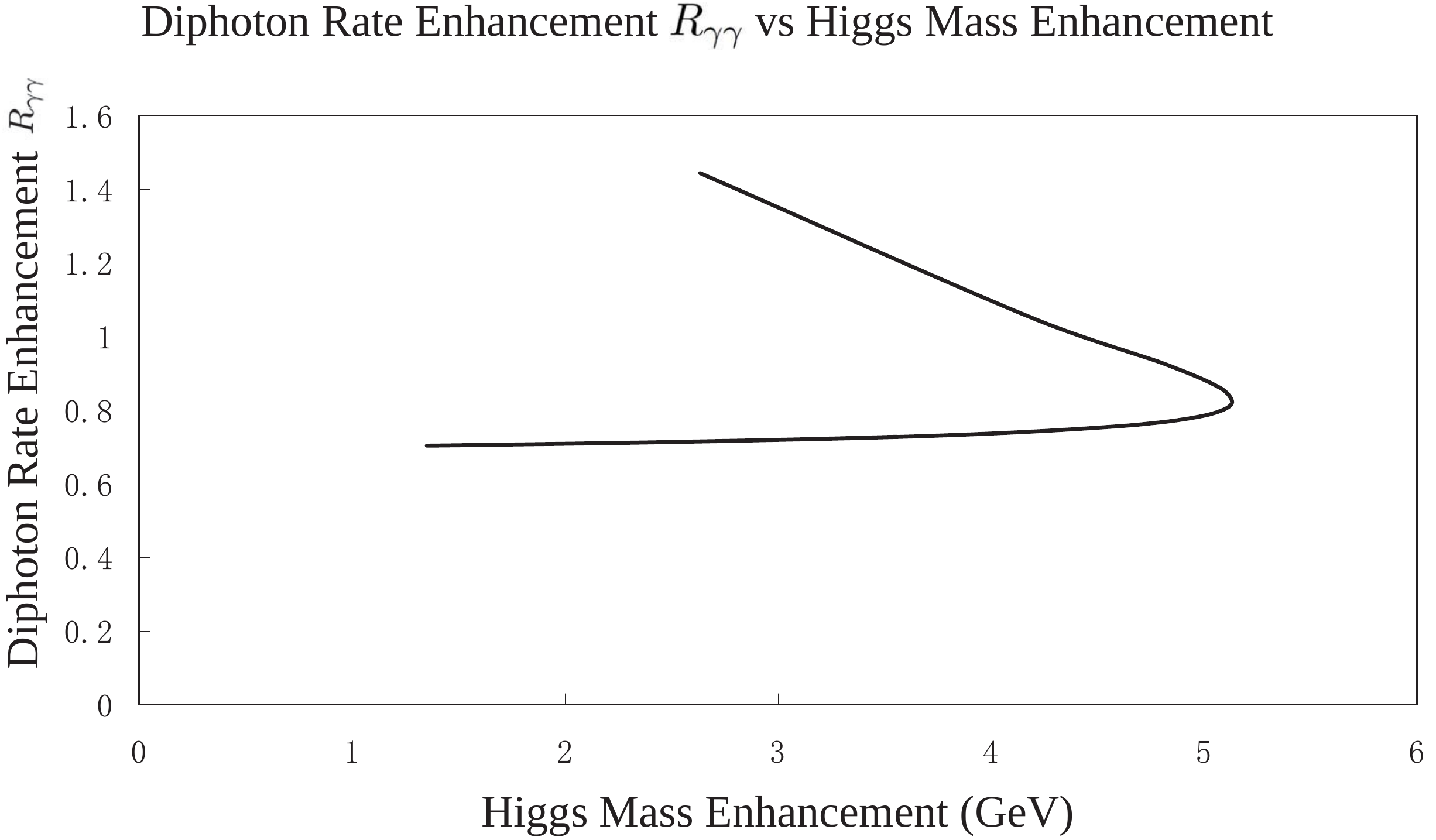}
\includegraphics[scale=0.33]{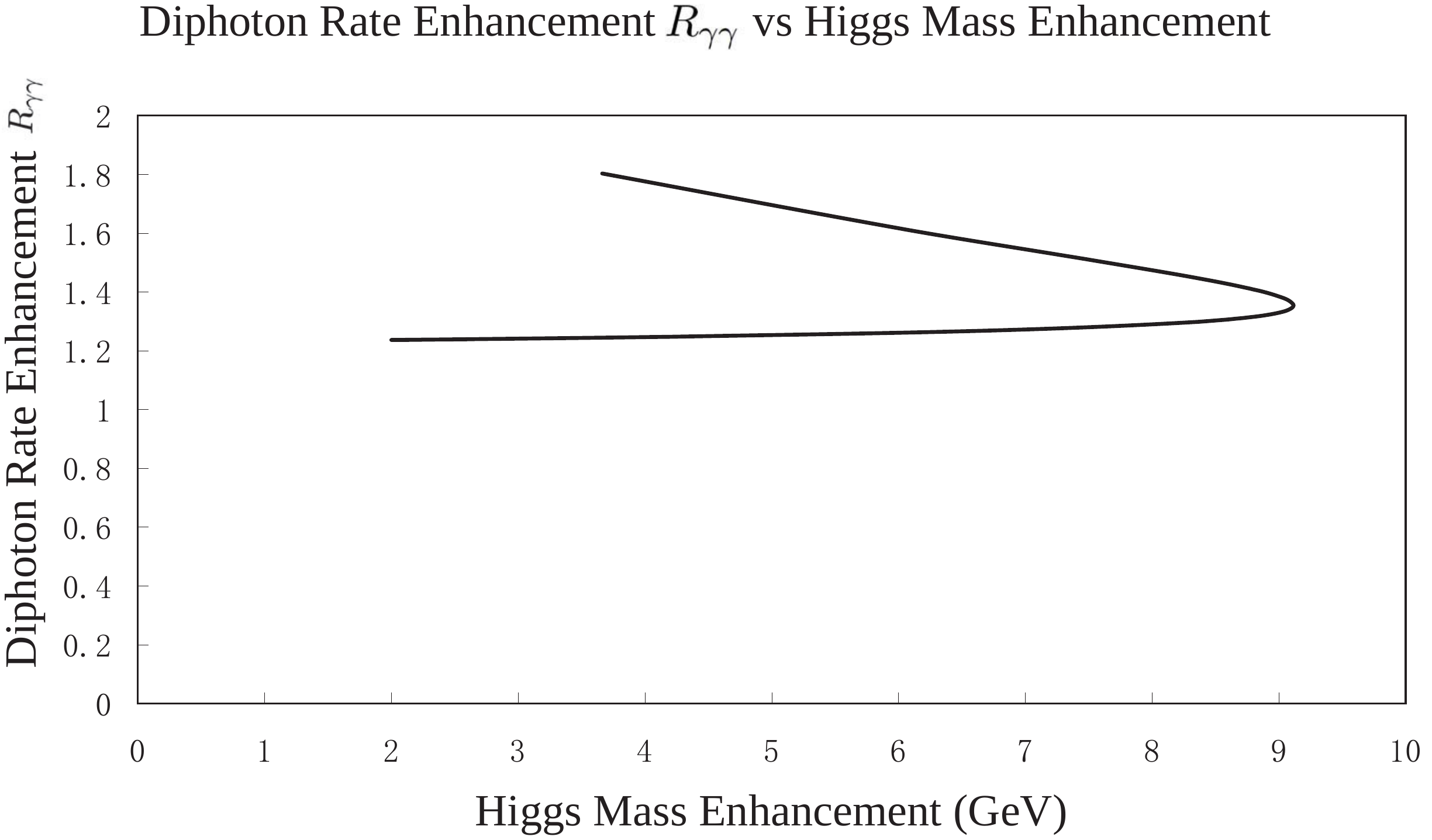}
\caption{ Left panel: A display of the correlation between the Higgs diphoton rate enhancement  and the Higgs mass enhancement
in the decoupled limit where $M_L=M_E=0$ as in Fig.~\ref{fig1}.
Right panel: A display of the correlation between the Higgs diphoton rate enhancement  and the Higgs mass enhancement
for the case when the vector masses are non-vanishing where $M_L=M_E=210\GeV$.
The two branches shown  in each  of the  two plots are due to the rise and fall
of the Higgs mass enhancement as exhibited in the lower panel of Fig.~\ref{fig1} and \ref{fig3}.}
\label{fig2}
\end{center}
\end{figure}

Before carrying out the  numerical analysis let us summarize the results of the analysis given in
Section~\ref{Sec4} and Section~\ref{Sec5}.
In Section~\ref{Sec4}  the correction to the Higgs diphoton rate from the fermionic sector $r_f$ was
computed in  Eq.~\eqref{fermionCon},
and the correction from the bosonic sector $r_b$ was computed  in Eq.~\eqref{rb}, while
 total diphoton rate enhancement $R_{\gamma\gamma}$ is given in Eq.~\ref{RGG2}.
The Higgs boson mass enhancement from the exchange of the vector-like supermultiplets
is given in Section~\ref{Sec5} where  the bosonic contribution is given by
Eqs.~\eqref{Db'11}-\eqref{Db'12} and \eqref{Db''11}-\eqref{Db''12} while  the fermionic contribution
is given by Eqs.~\eqref{Df11nz}-\eqref{Df12nz}.
In this section, for the numerical analysis we impose the constraint that the masses of the new particles
be consistent with the experimental lower limits~\cite{Beringer:1900zz}.\\

First, we discuss the decoupled limit where $M_L=M_E=0$.
In this case, both the fermionic sector and the bosonic sector of the vector-like
supermultiplets are totally decoupled, and we label the two sectors as the $\hat{\tau}'$ and $\hat{\tau}''$ sectors,
where $\hat{\tau}'$ denotes contributions from both $\tau'$ and its super-partners $\tilde\tau'_{1,2}$, and similar for $\hat{\tau}''$.
Here we choose the following parameters:
$M_1=M_2=500\GeV,\,\mu=1\TeV,\,\tan\beta=1.4,\,\alpha=\beta-\pi/2,\, y=y'=1$ and $m_h^{\rm MSSM}=120\GeV$.
Using the above parameters and Eq.~\eqref{fermionCon} we find that the fermionic contribution
to the Higgs diphoton rate $r_f$ is roughly $-0.4$ in this case,
which is a large negative effect.
However, this is compensated by the  contribution from the bosonic sector and this contribution is displayed  in the
upper two panels of  Fig.~\ref{fig1}.
The  upper left panel displays
the diphoton rate enhancements from the exchange of $\tilde{\tau}'_{1,2}$ in the loop
versus  $A_{\tau'}$ while the  upper right panel displays the diphoton rate enhancement from the exchange of
$\tilde{\tau}''_{1,2}$ versus  $A_{\tau''}$.
As expected in each case we find that the contribution from scalar loops enhances the diphoton rate.
The total contribution arising from the sum of the fermionic and the bosonic sectors
will be given when we discuss Fig.~\ref{fig2}.  \\

\begin{figure}[t!]
\begin{center}
\includegraphics[scale=0.33]{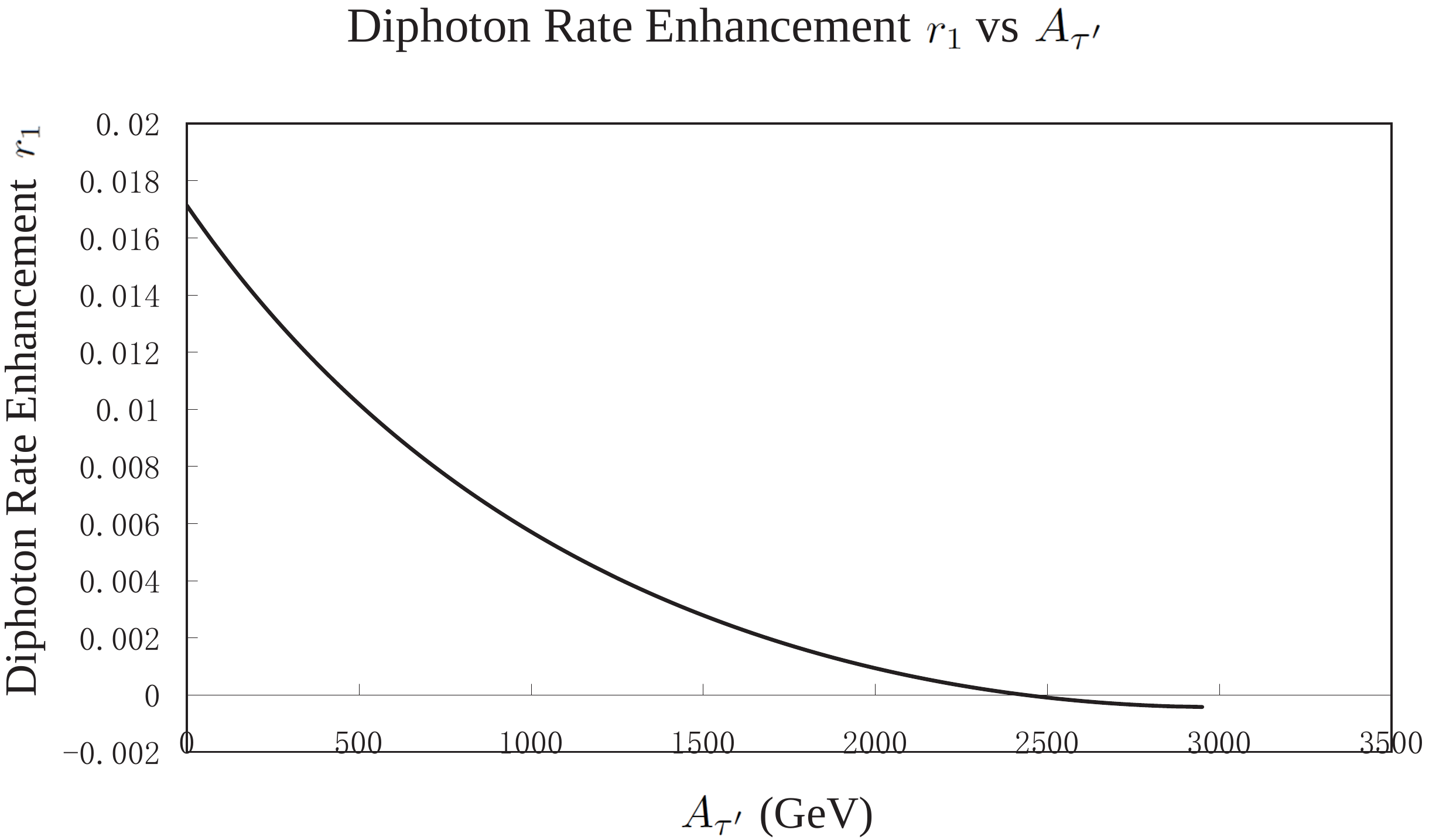}
\includegraphics[scale=0.33]{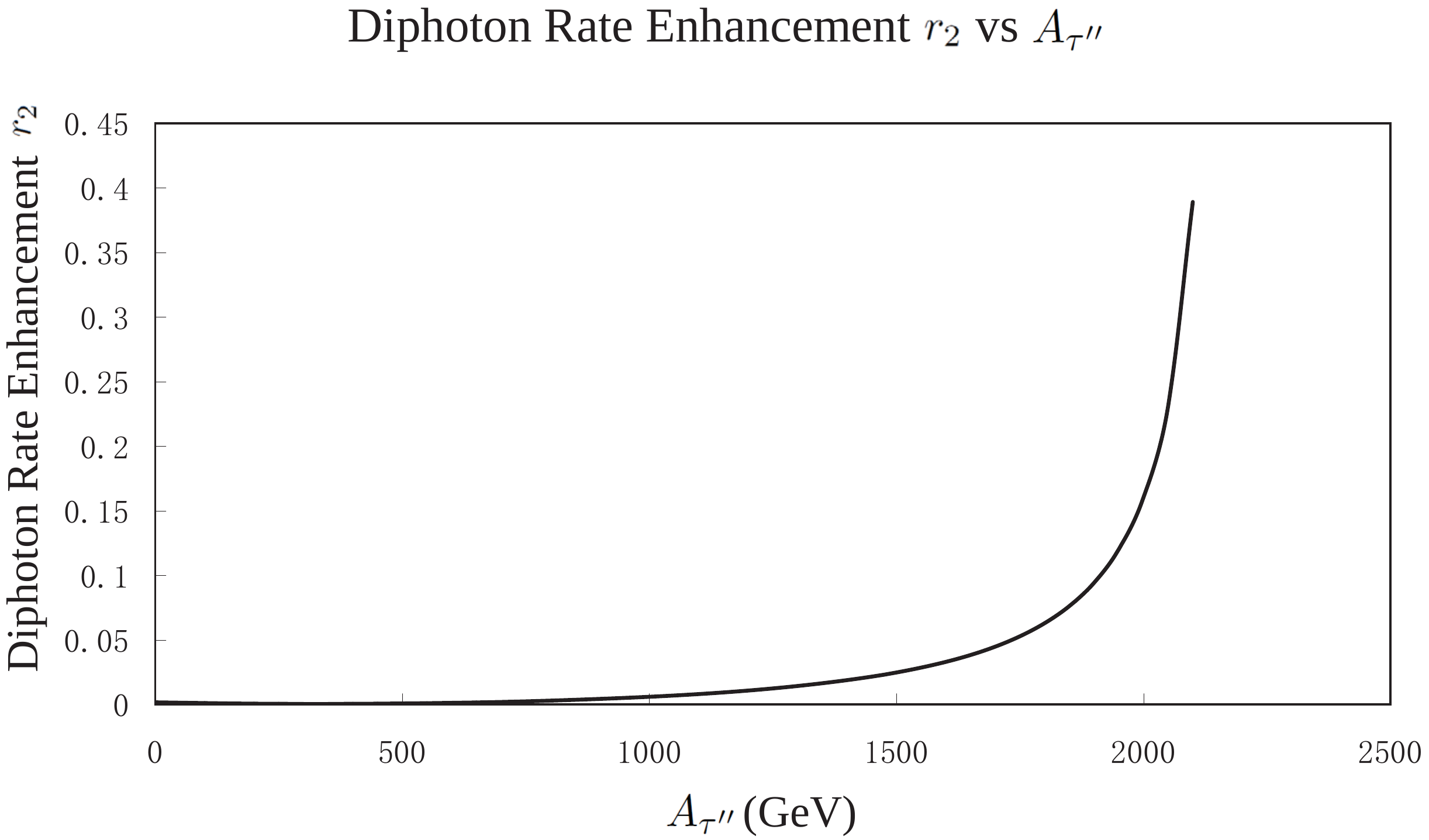}
\\~\\
\includegraphics[scale=0.33]{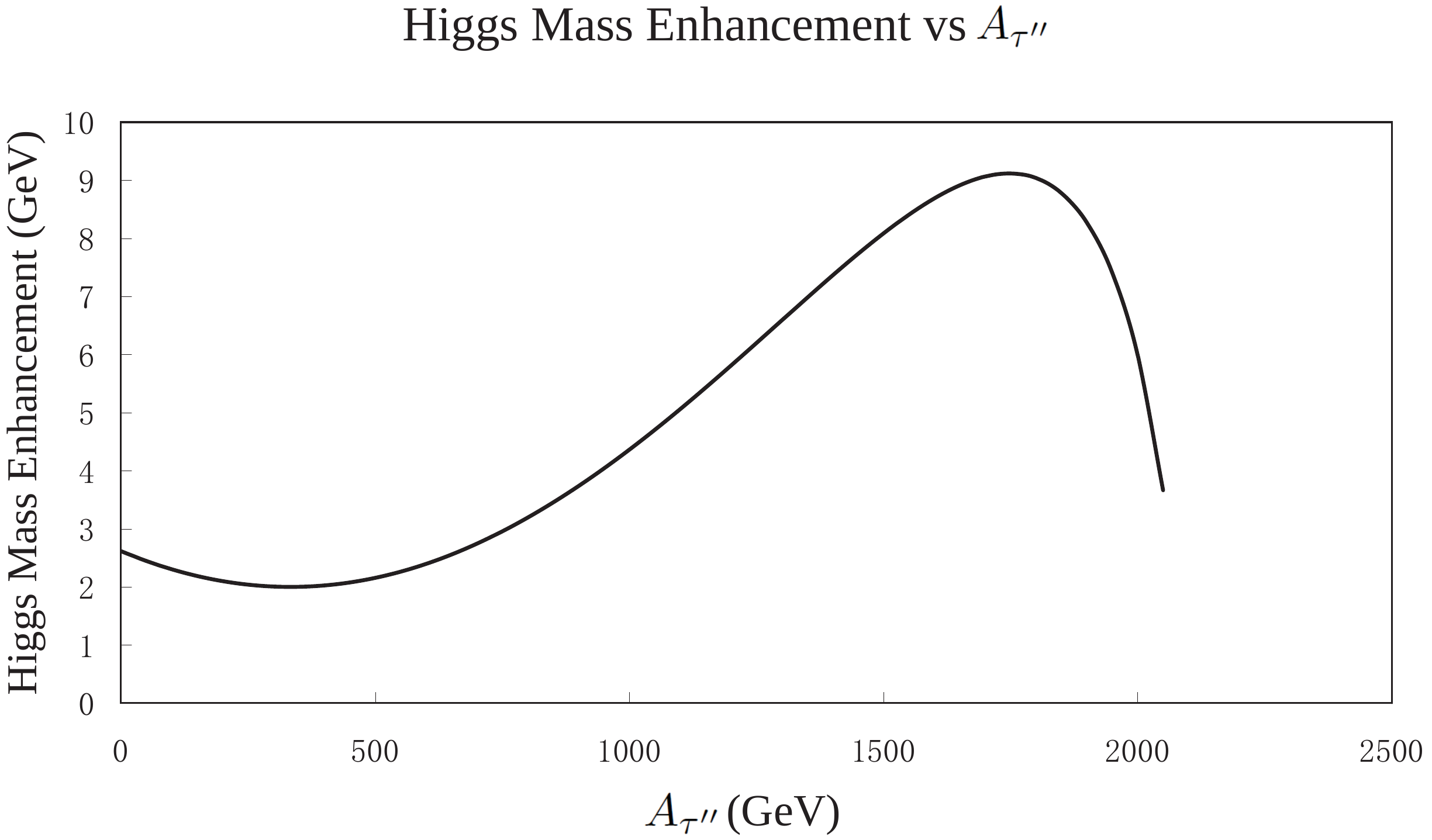}
\includegraphics[scale=0.33]{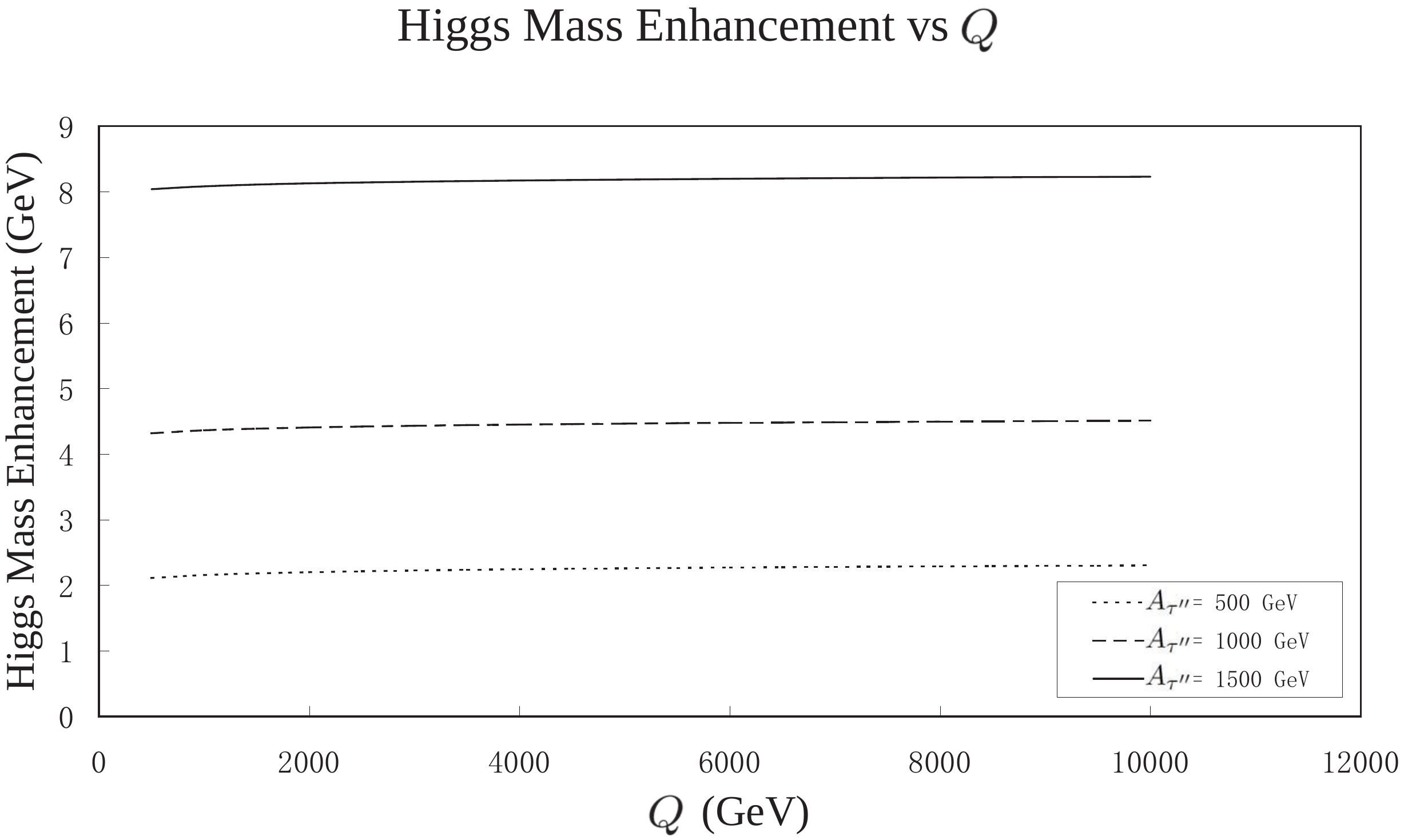}
\caption{
An analysis of the diphoton rate  enhancement (top panels) and enhancement of the Higgs boson mass
(bottom panels) for the case when the vector masses  are non-vanishing where $M_L=M_E=210\GeV$.
Left top: A plot of the Higgs diphoton rate enhancement $r_1$ (from $\tilde{\tau}'_{1,2}$) vs  $A_{\tau'}$;
Right top: A  plot of the Higgs diphoton rate enhancement $r_2$ (from $\tilde{\tau}''_{1,2}$) vs $A_{\tau''}$.
Left bottom: A plot of the total Higgs mass enhancement vs $A_{\tau''}$;
Right bottom:  A plot of the total Higgs mass enhancement vs the renormalization group scale $Q$.
The three horizontal lines correspond to values of $A_{\tilde \tau''}=500\GeV$ (bottom), $
1000\GeV$ (middle), $1500\GeV$ (top).
}
\label{fig3}
\end{center}
\end{figure}

An analysis of the enhancement of the Higgs boson mass in the decoupled case ($M_L=M_E=0$)
is given in the lower two panels of Fig.~\ref{fig1}.
The lower left panel of Fig.~\ref{fig1} gives a display of the Higgs mass enhancement from
the exchange of $\hat{\tau}'$ sector (including  bosonic and fermionic contributions) in the loop versus $A_{\tau'}$.
Here the contribution to the Higgs boson mass is rather modest not exceeding much
beyond 2 GeV over the entire range of $A_{\tau'}$.  A similar analysis for the mirror sector ($\hat{\tau}''$) is given in the
right panel of Fig.~\ref{fig1} where the Higgs boson mass enhancement is plotted against $A_{\tau''}$. Here
large contributions are seen to arise.
We turn now to a display of the combined diphoton rate from the fermionic and  the bosonic sector
versus the combined Higgs boson mass enhancement from the fermionic and bosonic sectors.
This analysis is presented in the left panel of
Fig.~\ref{fig2}  where we display  the total diphoton rate enhancement $R_{\gamma\gamma}$ as defined
in Eq.~\eqref{RGG2} versus  the total Higgs mass correction
(here we chose the maximum value for diphoton rate enhancement from $\tilde{\tau}'$ sector,
which corresponds to $A_{\tau'}=3100\GeV$).
While a simultaneous enhancement in both sector does occur, one finds in this case the sizes are rather modest,
e.g., one has a  3-4 GeV enhancement in the Higgs boson mass with a 30\% enhancement
in the diphoton rate at the same time.
\\

Next we  discuss the case when $M_L,M_E$ are non-vanishing.  Here we choose the following parameters:
$M_L=M_E=210\GeV,\,M_1=M_2=600\GeV,\,Q=\mu=1\TeV,\,\tan\beta=3,\,\alpha=\beta-\pi/2,\,y=y'=1$,
and $m_h^{\rm MSSM}=120\GeV$.
This time, the  contribution to the diphoton rate from the fermionic sector is positive and gives
 $r_f \approx +0.1$ on using Eq.~\eqref{fermionCon}.
The bosonic contribution is exhibited in the upper two panels of Fig.~\ref{fig3},
where the upper left panel displays the contribution from the exchange of
$\tilde{\tau}'_{1,2}$ in the loop
versus $A_{\tau'}$  while  the upper right panel  displays the contribution from  the exchange of
$\tilde{\tau}''_{1,2}$ in the loop versus  $A_{\tau''}$. Here essentially all of the
bosonic sector enhancement comes from the $\tau''$ sector.
\\

\begin{figure}[t!]
\begin{center}
\includegraphics[scale=0.32]{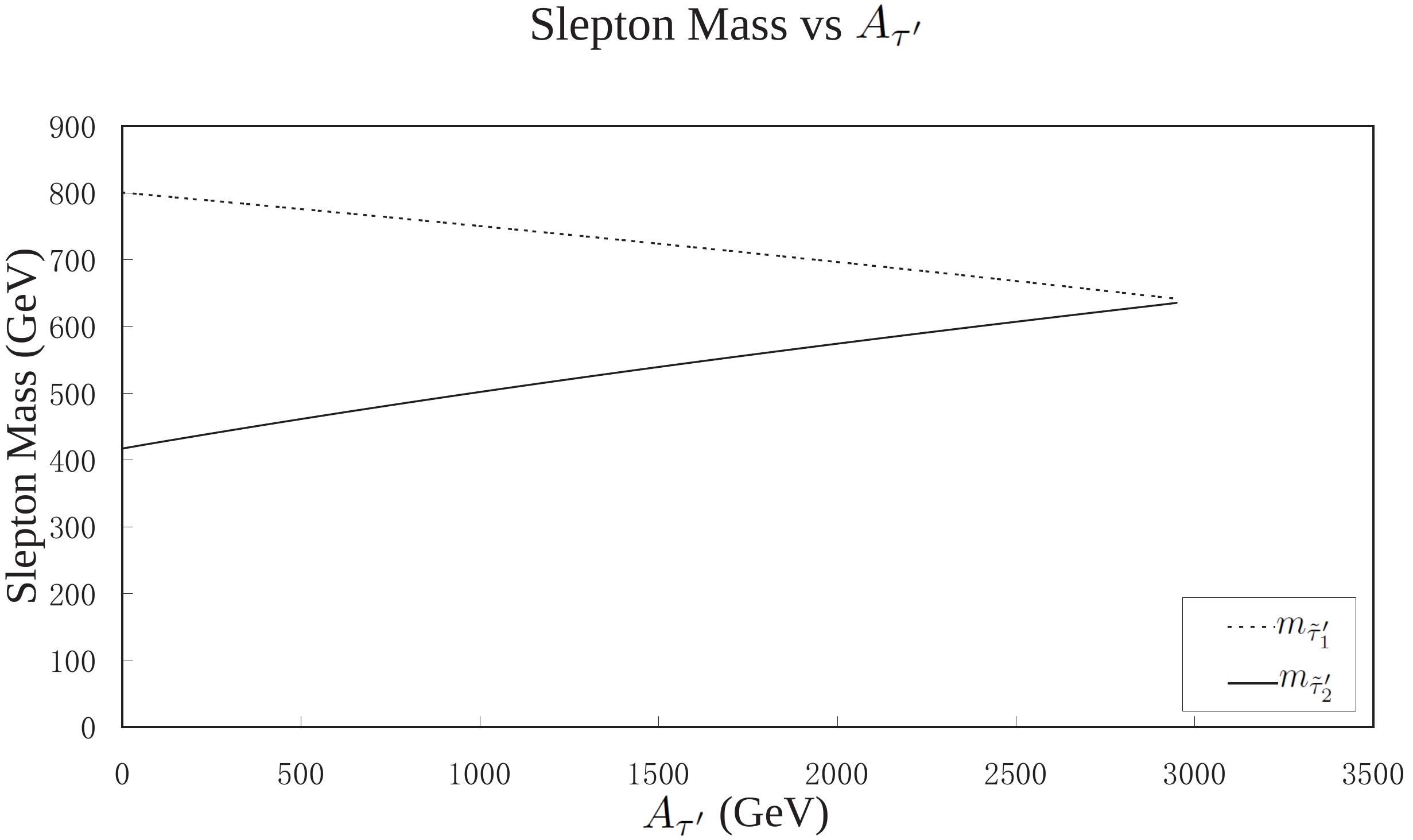}
\includegraphics[scale=0.32]{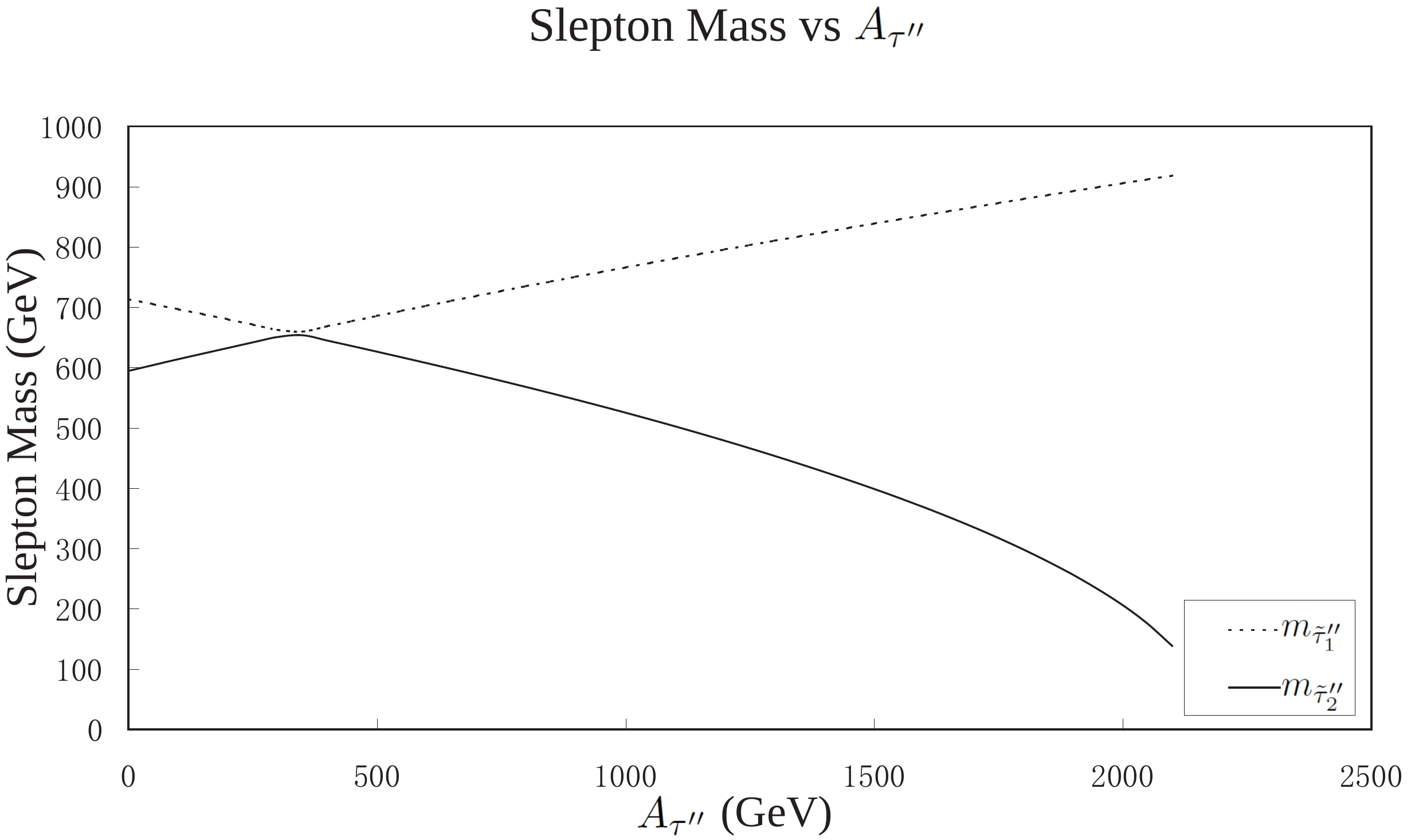}
\caption{A display of the slepton masses versus the trilinear couplings in the case $M_L=M_E=210\GeV$.
Left panel: A plot of the $\tilde{\tau}'_{1,2}$ masses vs $A_{\tau'}$. Right panel: A plot of
$\tilde{\tau}''_{1,2}$ masses vs $A_{\tau''}$.  We note that the slepton masses over most
of the parameter space lie significantly above the experimental lower limits~\cite{Beringer:1900zz}.}
\label{fig4}
\end{center}
\end{figure}

In the lower left panel of Fig.~\ref{fig3}
we display the {\em total} Higgs mass enhancements (adding up both the bosonic and fermionic contributions) versus $A_{\tau''}$,
where we choose $A_{\tau'}=1000\GeV$.
Similar to the diphoton enhancement,
the major contribution to the Higgs boson mass enhancement is also from the exchange of $\tilde{\tau}''_{1,2}$.
In the lower right panel of Fig.~\ref{fig3},
we display the total Higgs mass enhancement versus the renormalization group scale $Q$.
Again we choose $A_{\tau'}=1000\GeV$, and three specific values for $A_{\tau''}$
which correspond to three different values of the Higgs mass enhancement,
are chosen as shown in the plot.
The values for the scale $Q$ cover a large range from 500~GeV to 10~TeV,
and we see three almost straight  horizontal lines for the Higgs mass enhancement as a function of $Q$.
This plot shows the Higgs mass enhancement has almost no dependence on the scale $Q$,
which verifies that our approximation in computing the bosonic contribution to the Higgs mass is valid.
Combining the diphoton rate from both the bosonic and the fermionic sectors of the vector-like supermultiplets,
we display in the right panel of Fig.~\ref{fig2}
the total diphoton rate enhancement $R_{\gamma\gamma}$ versus the total Higgs mass correction
(where again we fix the contribution from $\tilde{\tau}'_{1,2}$ choosing $A_{\tau'}= 1000\GeV$).
Here  we find that including the vector masses, one can easily achieve
a diphoton rate enhancement as well as a Higgs mass enhancement of substantial size.
In Fig.~\ref{fig4} we give a display of the slepton masses. Here one finds that the slepton
masses from the new sector are typically in the few hundred GeV range except near the end points and lie substantially
above the experimental lower limits~\cite{Beringer:1900zz}. These mass ranges are consistent with
the electroweak constraints which  have been discussed in a number of
works~\cite{Cynolter:2008ea,Martin:2009bg,Martin:2010dc,ArkaniHamed:2012kq,Kearney:2012zi,Joglekar:2012vc}.
\\

Finally, we comment on the vacuum stability constraints.  These constraints on the $\hat\tau'$ and
$\hat\tau''$ sectors are similar to those discussed for the stau sector of MSSM and arise from the left-right mixing of the
staus~\cite{Hisano:2010re,Kitahara:2012pb,Carena:2012mw}.  The mixings lead to a cubic term
 in the Higgs potential expanded around
the electroweak symmetry breaking vacuum  which is of type $- y \mu h \tilde \tau'_L \tilde \tau'_R$ and
$-y' \mu h \tilde \tau''_L \tilde\tau''_R$. Such terms can generate global minima  in some cases.
The parameter that controls the instability is $\mu\tan\beta$. Without going into details because of the
smallness of  $\mu\tan\beta$ for the analysis given in Figs.~\ref{fig1}-\ref{fig4}
the solutions we present are consistent with the vacuum stability constraints.

\section{Conclusion}\label{Sec7}

In this work we consider an extension of MSSM with vector-like leptonic supermultiplets
and its possible implications to the Higgs diphoton rate and to the Higgs boson mass.
Specifically we compute one-loop corrections to the diphoton rate of the Higgs boson via
the exchange of the new leptons and their super-partners as well as their mirrors.
A similar analysis is carried out for the Higgs boson mass where
we compute corrections to its mass using the renormalization group improved Coleman-Weinberg
effective potential with contributions arising
also from these new particles.
It is found that an enhancement of the diphoton rate as large as  1.8
can occur and simultaneously a positive correction of 4-10 GeV to the
Higgs boson mass can also be obtained due to the exchange of the vector-like supermultiplets.
A correction of this size can have a significant effect
in relieving the  constraint on the weak scale supersymmetry.
\\

 In the supergravity unified model with universal boundary conditions at the GUT scale, one finds
that for a  Higgs mass in the 125-126 GeV region, the squark masses are rather heavy (see Fig.~1 of~\cite{Akula:2011aa})
and would be difficult to access at the LHC. However, a 5-10 GeV contribution to the Higgs mass from the new sector
would  put the MSSM component of the Higgs mass in the 116-120 GeV range which
allows a significantly lowering of the universal scalar mass (see Fig.~1 of~\cite{Akula:2011aa}).
Thus a Higgs mass correction of the size discussed in this work not only gives a significant correction to the
diphoton rate but also lowers the scale of supersymmetry, making sparticles more accessible in the next round of experiments at the LHC~\cite{Baer:2009dn}.
We also note that in the right panel of Fig.~\ref{fig4}
one finds that one of the scalar mass eigenvalue can lie close to the current experimental lower limit
and thus such states could be accessible at the  LHC and at the ILC.\\

The vector-like leptons can be produced at the LHC via processes such as
$pp\to Z \to \tau'\bar\tau'$.
The charged vector-like leptons will likely decay inside the detector via their gauge interactions
similar to any heavy lepton, e.g., $\tau' \to \tau \nu_{\tau'}\bar\nu_{\tau}$
with the subsequent decay of the $\nu_{\tau'}$. The decay of $\nu_{\tau'}$
would depend on mixings and is model dependent but in the end it could produce $l^+l^- \nu_{\tau}$.
In this case we  have as many as three charged leptons and missing $E_T$.
However, an accurate
analysis of the background is needed to quantify the size of the signal which is outside the scope of
this work.  Of course the best chance of seeing these particles
would be at the ILC through the process $e^+e^- \to Z \to \tau'\bar \tau'$ if sufficient center of mass
energy can be managed.\\

\noindent
{\it Acknowledgments:}
WZF is grateful to HaiPeng An and Tao Liu for helpful discussions.
The work of PN is supported in part by the U.S. National Science Foundation (NSF) grants
PHY-0757959 and  PHY-070467. WZF is supported by funds from The Hong Kong University of Science and Technology.

\appendix

\section{Loop functions}\label{AppA}
The loop functions $A_1(x), A_{\frac{1}{2}}(x)$ and $A_0(x)$ that appear in Section~\ref{Sec2} are defined by
\begin{align}
A_1(\tau)&=  -[2+ 3\tau + 3\tau(2-\tau) f(\tau)]\,,\\
A_{\frac{1}{2}}(\tau)&= 2\tau[1+ (1-\tau) f(\tau)]\,,\\
A_0(\tau)&= -\tau[1-\tau f(\tau)]\,.
\end{align}
Here the function $f(\tau)$ is defined by
\begin{equation}
f(\tau)=
\left\{
\begin{aligned} & \Big( \arcsin \frac{1}{\sqrt \tau} \Big)^{2}\,,\qquad\tau\geq1\,;\\
 & -\frac{1}{4}\Big[\ln\frac{\eta_{+}}{\eta_{-}}-i\pi\Big]^{2}\,,\qquad\tau<1\,.
\end{aligned}
\right.
\end{equation}
where $\eta_{\pm}\equiv (1 \pm \sqrt{1-\tau})$ and
$\tau = 4 m^2 \big/ m_h^2$ for a particle running in the loop with mass $m$.
For the case when $\tau \gg 1$ one has
\beqn
f(\tau)\to \frac{1}{\tau}(1+ \frac{1}{3\tau} + \frac{3}{ 20 \tau^2} +
 \cdots)\,,
\eeqn
and in this limit $A_1\to -7,\, A_{\frac{1}{2}}\to 4/3,\, A_0\to 1/3$.

\section{Loop corrections to the Higgs boson mass}\label{AppB}

In this Appendix we give details of the computation of corrections to the Higgs boson mass-squared
matrix arising from radiative corrections to the Higgs boson potential.
The Higgs potential is given by
\beqn
V(H_u,H_d)=V_0+\Delta V\,,
\eeqn
where $V_0$ is renormalization group improved tree-level potential and $\Delta V$ is the
loop correction.  For the case of two Higgs doublets in MSSM, including soft terms
the Higgs potential $V$ is given by
\begin{align}
V_0
&= \overline{m}_{H_u}^2|H_u|^2 + \overline{m}_{H_d}^2 |H_d|^2 +(|B\mu|^2 H_u.H_d + h.c.)
+\frac{(g_2^2-g_1^2)}{4}|H_u|^2|H_d|^2\nonumber\\
&+\frac{(g_2^2+g_1^2)}{8}|H_u|^4+
\frac{(g_2^2+g_1^2)}{8}|H_d|^4
-\frac{g_2^2}{2}|H_u.H_d|^2\,,
\end{align}
where $\overline{m}_{H_u}^2=M_{H_u}^2+|\mu|^2,\,\overline{m}_{H_d}^2=M_{H_d}^2+|\mu|^2$,
and $M_{H_{u,d}}$ and $B$ are the soft parameters.
The correction $\Delta V$ to the effective potential at the one loop level is given by~\cite{Coleman:1973jx,Arnowitt:1992qp}
\beqn
\Delta V=\frac{1}{64\pi^2}
Str \Big[ M^4_i(H_u,H_d)
\Big( \ln \frac{M^2_i(H_u,H_d)}{Q^2}-\frac{3}{2}\Big) \Big]\,,
\label{CWpoten}
\eeqn
where $M_i$ is the mass eigenvalue of the particle being exchanged,
$Str$ stands for the sum  $\sum_i c_{i} (2J_i+1)(-1)^{2J_i}$, $c_i(2J_i+1)$ counts the degrees of freedom, and the sum runs over
all the particles $i$ bosonic and fermionic being exchange in the loop.
Thus to construct the mass-squared matrix of the Higgs scalars
we need to compute the quantity
\beqn
(M_H)^2_{\alpha\beta}= \frac{\partial^2 V}{\partial v_{\alpha}\partial v_{\beta}}
=(M_H^{2})^0_{\alpha\beta}+ \Delta M^2_{H \alpha\beta}\,,
\eeqn
where $(\alpha, \beta)=(1,2)$ and $v_1\equiv v_d, ~v_2\equiv v_u$;
$(M_H^2)_{\alpha\beta}^{0}$ is the contribution from $V_0$ and
$\Delta M^2_{H \alpha\beta}$ is the contribution from $\Delta V$.
$\Delta M^2_{H \alpha\beta}$ is given by
\beqn
\Delta M^2_{H \alpha\beta}=
\frac{1}{32\pi^2}
Str \Big[ \frac{\partial M_i^2}{\partial v_{\alpha}}\frac{\partial M_i^2}{\partial v_{\beta}}
\ln \frac{M_i^2}{Q^2}+M_i^2 \frac{\partial^2 M_i^2}{\partial v_{\alpha}\partial v_{\beta}}
\Big( \ln\frac{M_i^2}{Q^2}-1\Big) \Big]\,.
\eeqn

In the analysis of corrections to the Higgs boson mass   the variations with respect to
the fields $v_u, v_d$ play an important role. Thus variations with respect to $v_u$ and
$v_d$ give the following two constraints
\begin{align}
\overline{m}_{H_d}^2+\frac{g_2^2+g_1^2}{8}(v_d^2-v_u^2)+|B\mu|^2 \tan\beta
+ \frac{1}{v_d}\frac{\partial \Delta V}{\partial v_d}&=0\,, \label{c1}\\
\overline{m}_{H_u}^2-\frac{g_2^2+g_1^2}{8}(v_d^2-v_u^2)+|B\mu|^2 \cot\beta
+ \frac{1}{v_u}\frac{\partial \Delta V}{\partial v_u}&=0\,.\label{c2}
\end{align}
In the computation of the Higgs boson mass-squared matrix it is found convenient to
eliminate $\overline{m}_{H_u}^2$ and $\overline{m}_{H_d}^2$ using the constraints of Eqs.~\eqref{c1} and \eqref{c2}.
This allows us to write
\beqn
M_H^2=
\left(\begin{matrix} M_Z^2c_{\beta}^2+M_A^2s_{\beta}^2+\Delta_{11} &
-(M_Z^2+M_A^2)s_{\beta}c_{\beta}+\Delta_{12} \cr
-(M_Z^2+M_A^2)s_{\beta}c_{\beta}+\Delta_{12} &
M_Z^2s_{\beta}^2+M_A^2c_{\beta}^2+\Delta_{22} \end{matrix}\right)\,,
\eeqn
where $M_Z^2= \frac{1}{4} (g_1^2+ g_2^{2})(v_u^2+ v_d^2)$ and $M_A^2= - 2|B\mu|^2\big/ \sin(2\beta)+\cdots$
and $\Delta_{ij}$ are now given by~\cite{Ibrahim:2000qj,Ibrahim:2002zk}:
\begin{align}
\Delta_{11} &= \Big(-\frac{1}{v_d} \frac{\partial}{\partial v_d} + \frac{\partial^2}{\partial v_d^2}\Big) \Delta V\,, \label{D11}\\
\Delta_{22} &= \Big(-\frac{1}{v_u} \frac{\partial}{\partial v_u} + \frac{\partial^2}{\partial v_u^2}\Big) \Delta V\,, \label{D22}\\
\Delta_{12} &=  \frac{\partial^2}{\partial v_u \partial v_d} \Delta V \label{D12} \,.
\end{align}
Evaluations of $\Delta_{ij}$ for the vector-like leptonic supermultiplet are given in Section~\ref{Sec5}.


\end{document}